\documentclass[a4paper,showpacs,superscriptaddress,nofootinbib,10pt]{revtex4}
\usepackage{amsmath,amssymb,bm,natbib}
\usepackage{epsfig}
\usepackage{epstopdf}
\usepackage{graphicx}
\usepackage{multirow}
\usepackage{slashed}
\makeatletter

\newcommand{\Rmnum}[1]{\expandafter\@slowromancap\romannumeral #1@}
\makeatother
\begin{document}
\def\bib{\bibitem}
\def\be{\begin{equation}}
\def\ee{\end{equation}}
\def\beq{\begin{equation}}
\def\eeq{\end{equation}}
\def\beqar{\begin{eqnarray}}
\def\eeqar{\end{eqnarray}}
\def\barr{\begin{array}}
\def\earr{\end{array}}
\def\lsim{\:\raisebox{-0.5ex}{$\stackrel{\textstyle<}{\sim}$}\:}
\def\gsim{\:\raisebox{-0.5ex}{$\stackrel{\textstyle>}{\sim}$}\:}
\def\tilh{\tilde{h}}
\def\and{\qquad {\rm and } \qquad}
\def\vev{{\it v.e.v. }}
\def\p{\partial}
\def\ga{\gamma^\mu}
\def\slp{p \hspace{-1ex}/}
\def\sleps{ \epsilon \hspace{-1ex}/}
\def\slk{k \hspace{-1ex}/}
\def\slq{q \hspace{-1ex}/\:}
\def\prl#1{Phys. Rev. Lett. {\bf #1}}
\def\prd#1{Phys. Rev. {\bf D#1}}
\def\plb#1{Phys. Lett. {\bf B#1}}
\def\npb#1{Nucl. Phys. {\bf B#1}}
\def\mpl#1{Mod. Phys. Lett. {\bf A#1}}
\def\ijmp#1{Int. J. Mod. Phys. {\bf A#1}}
\def\zp#1{Z. Phys. {\bf C#1}}
\def\etal{ {\it et al.} }
\def\ie{ {\it i.e.} }
\def\eg{ {\it e.g.} }
\def\sbar{ \overline{s} }
\def\thmin{\theta_0}
\def\cmin{\cos \theta_0}
\def\pepebar{P_{e} P_{\overline{e}}}
\def\eebar{$e^+e^-~$}
\def\eegz{$e^+e^- \to \gamma Z$}
\def\ggz{$\gamma\gamma Z~$}
\def\gzz{$\gamma ZZ~$}

\title{ Probing the indefinite CP nature of the Higgs
Boson through decay distributions in the process $e^+e^-\to t\bar{t}\Phi$
}
\author{{\bf B. Ananthanarayan}}
\affiliation{
Centre for High Energy Physics, 
Indian Institute of Science, 
Bangalore 560 012, India} 
\author{{\bf Sumit K. Garg}}
\affiliation{
Department of Physics and IPAP, Yonsei University,
 Seoul 120-749, Korea}
\author{{\bf Jayita Lahiri}}
\affiliation{
Centre for High Energy Physics, 
Indian Institute of Science, 
Bangalore 560 012, India} 

\author{{\bf P. Poulose}}

\affiliation{ 
Department of Physics,
Indian Institute of Technology Guwahati,
Guwahati 781 039, India}


\begin{abstract}
The recently discovered scalar resonance at the LHC is now almost confirmed to be a Higgs 
Boson, whose CP properties are yet to be established.  At the ILC with and without 
polarized beams, it may be possible to probe these properties at high precision.  
In this work, we study the possibility of probing departures from the pure CP-even case, 
by using the decay distributions in the process $e^+ e^- \to t \bar{t} \Phi$, with $\Phi$
mainly decaying into a $b\bar b$ pair.  
We have compared the case of a minimal extension of 
the SM case (Model I) with an additional pseudoscalar degree of freedom, with a more 
realistic case namely the CP-violating Two-Higgs
Doublet Model (Model II) that permits a more general description of the couplings.
We have considered the ILC with $\sqrt{s}=800$\,GeV and integrated luminosity of $300\, 
{\rm fb}^{-1}$.  Our main findings are that even in the case of small departures 
from the CP-even case, the decay distributions are sensitive to 
the presence of a CP-odd component in Model II, while it is difficult to probe these 
departures in Model I unless the pseudoscalar component is very large. Noting that the 
proposed degrees
of beam polarization increases the statistics, the process demonstrates the
effective role of beam polarization in studies beyond the Standard Model. Further, 
our study shows that an indefinite CP Higgs would be a sensitive laboratory to physics
beyond the SM.

\end{abstract}

\pacs{13.66.-a, 12.60.-i, 13.88.+e, 12.60.Fr}
\maketitle

\section{Introduction}
Recently the Large Hadron Collider (LHC) discovered a new particle resonance 
that weighs about  125\,GeV$/c^2$, see Refs.~\cite{cms,atlas},
which may indeed be the long sought after Higgs Boson. 
While its presence is not in doubt, the properties 
of this particle are yet to be completely determined. 
What is clear is that it is electrically neutral, 
and that it is not a vector resonance, and almost certainly
has spin-0. 
Whether or not it is the Standard Model (SM) Higgs Boson will be decided by the spin and partiy properties of the particle, 
and also its couplings with other particles. 
The LHC being a hadronic machine may not be able to exhaustively
study the properties of this particle. 
The proposed International Linear Collider (ILC) ~\cite{ILC1,ILC2}, which
is an $e^{+}-e^{-}$ collider, will carry out precision experiments
on SM particles and establish their properties
including that of the purported Higgs Boson.  
It has been pointed out, see Ref.~\cite{polarizationreview} that
beam polarization could significantly enhance the sensitivity of the
machine to probe beyond the SM signals.  

The Standard Higgs mechanism employed by the SM as a solution 
to the electroweak symmetry
breaking (EWSB) introduces one Higgs state (doublet of $SU(2)_L$), 
resulting in a physical CP-even scalar particle. 
However, many extensions of the EWSB mechanism like the Supersymmetric
(SUSY) extensions or the Two-Higgs Doublet Model (2HDM) predict 
more than one physical scalar particle. In CP-conserving models, 
these states are either CP-even or CP-odd. 
While it is too early to identify the CP nature of the new resonance at 
125\,GeV$/c^2$, it is almost certainly not a purely CP-odd state. 
This follows from the fact that one of the discovery channels
involve $ZZ\Phi$ coupling, which would have been absent if $\Phi$ was purely CP-odd. 
This was also indicated from the recent data analysis results of  both ATLAS\cite{Moroind1,Moroind2,Moroind3} 
and CMS\cite{Moroind4,Moroind5}, where they set an exclusion limit of around $3\sigma$ for pure CP-odd ($J^P = 0^-$) state. 
However a CP-mixed state can fit in very well with the data. 

CP-mixed states are possible in CP-violating
versions of the SUSY models \cite{CPsuperH}, as well as in the general 2HDM with CP 
violation in the Higgs sector \cite{2HDMcpv}.  
Indeed, there had been many interesting studies on 
CP-violating Higgs sector within different versions of the SUSY models \cite{MSSMcpv}, 
and in other models. Some of the recent studies along these 
lines in the light of new LHC discovery may be found in Ref.~\cite{MSSMcpvPP} and references therein.  Being heaviest among 
the SM particles, the top quark coupling to the Higgs Boson is the 
strongest, and therefore most promising to study the CP nature of the resonance. 
A recent study ~\cite{DDGMR} has pointed out that the ILC is an ideal setting to 
probe the CP nature of the Higgs Boson in the process 
\begin{equation}\label{ourprocess}
e^+e^-\to t \bar{t}\Phi.
\end{equation}
Here the deviation of the Higgs coupling to the top quark was 
parametrized by considering a CP-mixed Higgs state. 
The scalar and pseudo-scalar parts of such a CP-mixed Higgs Boson 
will couple differently to different polarization combinations 
of the top quark and top antiquark. A measurement of top 
quark polarization (and/or polarization asymmetry) 
could therefore probe the CP-nature of the 
particle.  More recently, in Ref.~\cite{GHMRS}, it
was shown that a combined use of total cross section and its
energy dependence, the polarization asymmetry of the top quark
and the up-down asymmetry of the antitop with respect to
the top-electron plane can significantly help in determining
the CP properties in the event of CP conservation and in
that of mixing in the case of CP violation.  
The properties in the decay to $\tau$ lepton
pairs has also been considered, see Ref.~\cite{BBS},
in the process $e^+e^-\to Z \Phi$. Related papers are Refs.~\cite{HJ,RR}.

It is worth pointing out that, quite independently of the considerations
of the CP properties of the Higgs Boson, there have been several 
studies of process in Eq. (\ref{ourprocess})
in the context of the measurement of the top quark
Yukawa coupling to the Higgs. For some
early work, see, e.g. Ref.~\cite{AG1,AG2}
and references therein.  More recently the process
has attracted renewed attention:  in Refs.~\cite{KS1,KS2} the
size of the signal and backgrounds when various decays
are considered has been studied, while the issue of
the process at a centre of mass energy of 500\,GeV 
due to the QCD enhancement of the crosssection near $t\bar t$ threshold 
has been considered in some detail in Ref.~\cite{YIUF,YITFKSY}, and finally the issue 
of a direct measurement using the semi-leptonic final state from the
decays of the $W$ arising from the decays of the top quarks
is considered in Ref.~\cite{TM}. 

Keeping in mind the above considerations, we now wish to study
the possibility of fingerprinting the departure from the CP-even case 
\cite{AguilarSaavedra} in decay distributions of the process in Eq. (\ref{ourprocess}),
which will necessarily require us to go beyond the analytical
approach.  
In our study we commit ourselves to two definite scenarios 
which we denote as Model I and Model II.  
Model I corresponds to the minimal extension of the SM with one additional pseudo-scalar 
degree of freedom, which mixes with the SM scalar to form the physical Higgs 
Boson~\cite{DDGMR}. This model is  characterized by one free 
parameter, which is denoted by $b$. Model II is a more realistic case similar to the 
CP-violating 2HDM model which has some 
essential features that make it quite different from Model I.
In particular, there is no theoretical constraint on parameters denoted by 
$a$, $b$ and $c$ (as described in the
Section~\ref{formalism}) and thus permits a more general discussion.  
However, we confine ourselves to some reasonable ranges for these parameters 
guided by the experimental indications that the resonance is close to a CP-even case.

In order to meet our objectives, considering that these are
not amenable to analytical methods, and must necessarily involve
numerical packages of great sophistication and complexity,
we have used the integrated Monte-Carlo and event 
generation package {\sc WHIZARD} \cite{WHIZARD} 
for our study.  The SM, as well 
as some of its popular extensions are
already implemented in this package.  
Further, any new model described through a Lagrangian can be 
incorporated  into this package through an interface 
\cite{Interface} generated using FeynRules \cite{FeynRules}. 
In particular, in our work we introduce decays for the top-quark 
which can be implemented in WHIZARD.  Our signal processes are
\begin{equation}\label{signalprocess}
e^+e^-\rightarrow t\bar t \Phi\rightarrow  W^+W^-b\bar b \Phi,\, t\bar{t} b\bar{b},
\end{equation}
where the first final state is the result of $t\bar t$ pair decay, keeping $\Phi$
fully reconstructed, while the second one considers the decay of $\Phi$ into $b\bar b$
pair. In contrast to the existing studies of the $t\bar t \Phi$ production, we
consider the effect of CP-violating Higgs Boson in the decay spectrum of both the top
quark as well as the Higgs boson itself, noting that the decay distributions are the
spin analysers of the parent particle.


The scheme of this paper is as follows.  In Section \ref{formalism}  we first
introduce and describe the basic structure of an indefinite CP Higgs
sector in the two scenarios mentioned above.  
In Section \ref{process} we describe the processes we consider.  
In Section \ref{results}
we present the results of our analysis.  In Section \ref{discussion}
we present a discussion
and our conclusions.  

\section{Formalism and Models}\label{formalism}
In this section we present the formalism we have adopted.
In the Standard Higgs mechanism with one  Higgs doublet 
acquiring vacuum expectation value (vev), 
there is only one CP-even physical scalar field. 
In the simplest extensions of this with an additional complex singlet or complex doublet,
there are CP-odd states along with one or more CP-even states. If CP symmetry is violated
in the Higgs sector, the physical Higgs states could be in a CP-mixed state. 
In this article we focus our attention on such a scenario, and its possible
implications on the $t\bar t \Phi$ production at the ILC. A study of this process will thus
give information on the CP nature of the Higgs Boson. 

\begin{figure}[t]
\begin{center}
\includegraphics[width=0.45\textwidth]{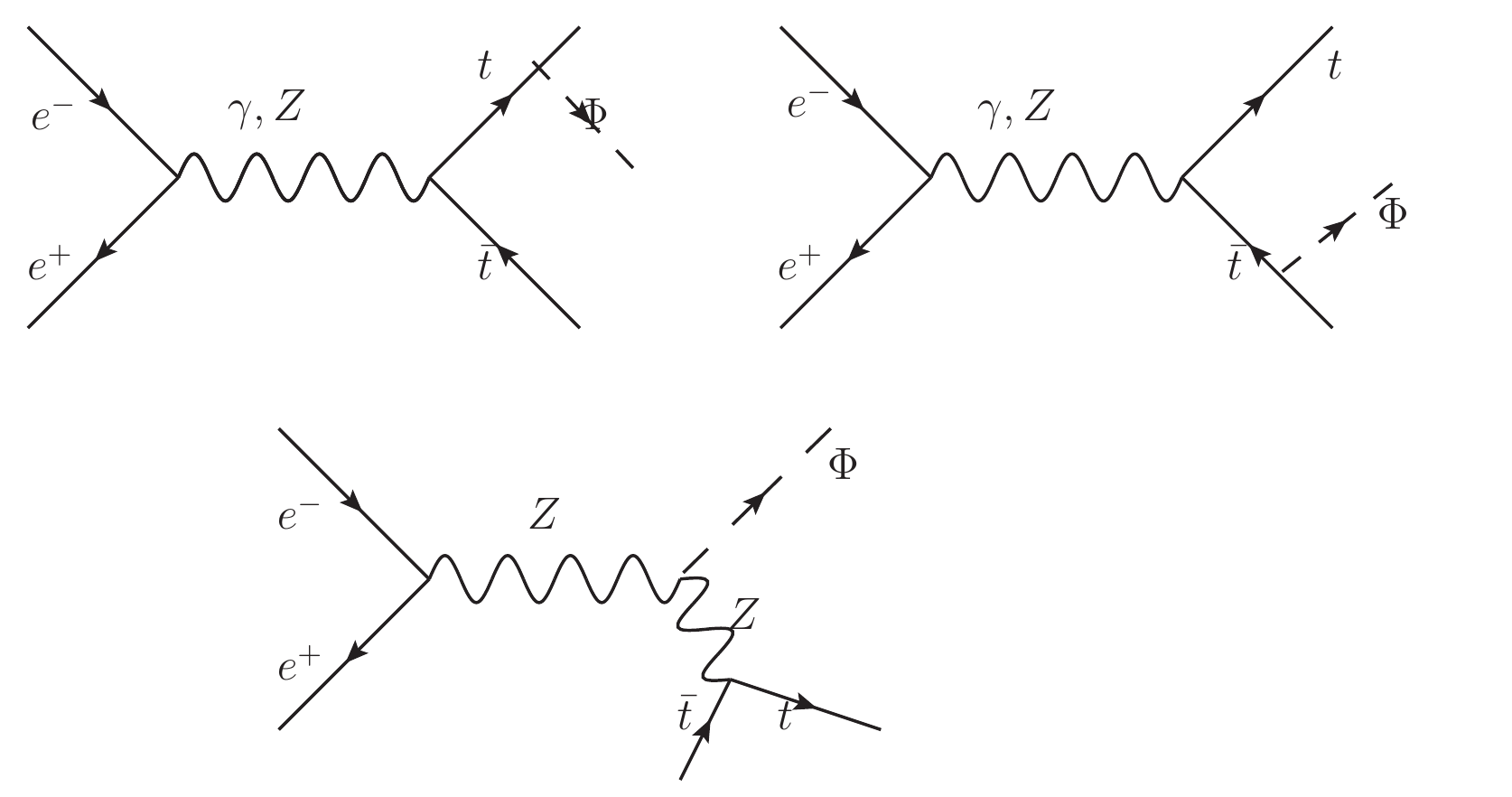}
\end{center}
\caption{Feynman diagrams contributing to  the process $e^{-}e^+ \rightarrow t\bar{t}\Phi$ in Standard Model. }
\label{fig:0}
\end{figure}

The process $e^+e^-\rightarrow t\bar t \Phi$ goes through the channels shown in
Fig.~\ref{fig:0}. The two channels with Higgs Boson radiating off the top quark or
antiquark dominate the cross section, with about a few percent contribution from the
third channel with the Higgs radiating off the $Z$.

With CP-mixed Higgs Boson, both the $t\bar t \Phi$ as well as $ZZ\Phi$ couplings take a form,
which may be parametrized as follows~\cite{AguilarSaavedra, DDGMR}
\begin{eqnarray}
g_{\Phi tt}&=&-i \frac{e}{s_W} \frac{m_t}{2 M_W} (a+i b \gamma_5) \nonumber \\
g^{\mu\nu}_{ZZ\Phi}&=&-ic \frac{e M_Z}{s_W c_W} g^{\mu\nu}.
\label{eq:generic_coupl_t}
\end{eqnarray}
In the above, 
$s_W\equiv \sin\theta_W(=\sqrt{1-c_W})$, where $\theta_W$ is the Weinberg angle.  
In the SM with only one scalar Higgs Boson ($h$), the parameters take values $a=1$, 
$b=0$ and $c=1$.

\subsection{Model I}

In a minimal extension of the SM case, one imagines the presence of an additional pseudo-scalar
degree of freedom $A$, which mixes with the scalar degree of freedom to produce a physical state: 
\begin{equation}
\Phi = a~h + b~A.
\end{equation}
We call this scenario as Model I in the rest of this article.
The parameters $a$ and $b$ represent the mixing, and are related to each
other by 
\begin{equation}
a^2+b^2=1.\label{model1constr}
\end{equation}
Since the SM gauge Boson, $Z$ does not couple to the pseudo-scalar degree
of freedom, we have $c=a$ in this scenario. The down-type quarks as well as the charged
leptons will also have the same coupling structure as that of the up-type quarks, so
that, for example, the $b$-quark couplings become
\begin{eqnarray}
g_{\Phi bb}&=&-i \frac{e}{s_W} \frac{m_b}{2 M_W} (a+i b \gamma_5).
\label{eq:generic_coupl_b}
\end{eqnarray}

\subsection{Model II}

While model I has the advantage in phenomenological analysis that there is only one free 
parameter, in most of the realistic cases of extensions beyond the SM the Higgs sector is 
more 
complex. For example, in the 2HDM and in the Minimal Supersymmetric Standard Model (MSSM) there are two Higgs 
doublet fields, leading to two neutral scalar Bosons and one pseudoscalar Boson in the 
physical spectrum in the CP-conserving case.  Denoting the gauge eigenstates of the
scalar fields as $\phi_1$ and $\phi_2$, and the mass eigenstates as $h$ and $H$, we can
write down the relation between them in terms of the mixing matrix as follows. 
\begin{equation}
\left(\begin{array}{c} H\\[1mm] h\end{array}\right)=
\left(\begin{array}{cc}\cos\alpha&\sin\alpha\\[1mm]
                  -\sin\alpha&\cos\alpha \end{array}\right)~
\left(\begin{array}{c}\phi_1\\[1mm] \phi_2\end{array}\right).
\end{equation}

In the CP-violating case, all the three degrees of freedom mix to give CP-mixed 
physical mass eigenstates as below.
\begin{equation}
\left(\begin{array}{c} \phi_1\\[1mm]  \phi_2\\[1mm] A\end{array}\right)=
{\cal O}_{3\times 3}~
\left(\begin{array}{c}H_1\\[1mm] H_2\\[1mm] H_3\end{array}\right),
\end{equation}
where $A$ is the pseudo scalar gauge eigenstate \cite{CPsuperH}.
This, in effect, removes the restricting relations between the parameters $a,~~b$ and $c$.
For ready reference we take the example of MSSM case (or 2HDM) with and without CP violation in the Higgs
sector, and list in Table~\ref{tab:mssm} the couplings of the Higgs Boson with the 
fermions and the gauge Bosons, where $\tan\beta$ is the ratio of the vev's of the two Higgs 
fields. Comparing Table~\ref{tab:mssm} with Eq.~\ref{eq:generic_coupl_t}, \ref{eq:generic_coupl_b}, we have 
\begin{eqnarray}
{\rm top~quark:}&&~~a_u=~{\cal O}_{2i}~/~\sin\beta,~~~~~~~b_u=-{\cal O}_{3i}~\cot\beta\nonumber\\
{\rm bottom~ quark / \tau -lepton:}&&~~a_d=~{\cal O}_{1i}~/~\cos\beta,~~~~~~~b_d=-{\cal O}_{3i}~\tan\beta\nonumber\\
{\rm gauge~ Bosons:}&&~~c=~{\cal O}_{1i}~\cos\beta+{\cal O}_{2i}~\sin\beta,
\label{eq:cpvmssm_coupl}
\end{eqnarray}
where we have introduced the subscripts $u$ and $d$ on the parameters $a$ and $b$ to
denote the up-type and down-type quarks, respectively.  The mixing matrix elements 
satisfy the normalization conditions:
\begin{equation}
{\cal O}_{1i}^2+{\cal O}_{2i}^2+{\cal O}_{3i}^2=1.
\end{equation}

\begin{table}
\begin{tabular}{|l|c|c|c|c||c|}
\hline
&&\multicolumn{3}{c||}{CP-conserving}&CP-violating\\ \cline{3-6}
&$g_{SM}$&$\Phi=$h&H&A&$H_i$\\ \cline{1-6}
&&&&&\\
$g_{\Phi tt}$&$-i \frac{e}{s_W} \frac{m_t}{2
M_W}$&$g_{SM}~\frac{\cos\alpha}{\sin\beta}$&$g_{SM}~\frac{\sin\alpha}{\sin\beta}$&
$g_{SM}~(-i)~\cot\beta~\gamma^5$&$g_{SM}~\left({\cal O}_{2i}~/~{\sin\beta}-i~{\cal
O}_{3i}~\cot\beta~\gamma^5\right)$\\[2mm]\cline{1-6}
&&&&&\\
$g_{\Phi bb}$&$-i \frac{e}{s_W} \frac{m_b}{2
M_W}$&$g_{SM}~\frac{-\sin\alpha}{\cos\beta}$&$g_{SM}~\frac{\cos\alpha}{\cos\beta}$&
$g_{SM}~(-i)~\tan\beta~\gamma^5$&$g_{SM}~\left({\cal
O}_{1i}~/~{\cos\beta}-i~{\cal O}_{3i}~\tan\beta~\gamma^5\right)$\\[2mm]\cline{1-6}
&&&&&\\
$g_{\Phi VV}$&$-i g_VM_V$&$g_{SM}~\sin(\beta-\alpha)$&$g_{SM}~\cos\left(\beta-\alpha\right)$&0
&$g_{SM}~\left({\cal O}_{1i}~ \cos\beta+{\cal O}_{2i}~\sin\beta\right)$\\
&&&&&\\
\hline
\end{tabular}
\caption{Couplings of the Higgs Bosons in the CP-conserving and CP-violating 2HDM. 
$V=W,~Z$, with $g_W=e/\sin\theta_W$, $g_Z=e/(\sin\theta_W \cos\theta_W)$.}
\label{tab:mssm}
\end{table}
We call this scenario as Model II in the rest of this article.
The lightest of the Higgs Bosons, $H_1$ will be assumed to be the discovered 125 GeV
resonance (denoted as $\Phi$) , while $H_2$ and $H_3$ are considered to be heavy
enough to be out of LHC range investigated so far.

\subsection{Features of Models I and II}

While the spin and parity measurements of the LHC resonance are not conclusive yet, LHC
reports that the new resonance is consistent with a $J^P=0^+$. It may be noted that these
analyses are done with the either or hypothesis. The scope of a mixed CP-state need a more
complex analysis, which may be beyond the capability of LHC at present. While in Model
I the parameters $a$ and $b$ are directly proportional to the scalar and pseudoscalar
components of the Higgs Boson, in Model II, even with 
small mixings, it is possible to have large changes in the couplings $a$ and $b$,
owing to the relations expressed in Eq.~\ref{eq:cpvmssm_coupl}. 
Taking the spin and parity
measurements at LHC seriously, one may consider the mixing matrix elements $O_{31}$ to be
small. Further, considering the large contribution of top quark loop to both the diphoton
decay process, as well as the gluon fusion production of the Higgs Boson, let us consider
the case where the coupling $a_u$ is close to 1. In Table~\ref{tab:2HDM_coupl} we
present a few possible sets of values of parameters along this line, corresponding to
pseudo scalar component of 1\%  ($O_{31}=0.10$) 
for two different $\tan\beta$ values of 2 and 20. For $a_u$ to be large, the scalar 
component should be mostly of $\phi_2$ type. We have considered the case with a small
admixture (about 5\%) of  $\phi_1$ (corresponding to $O_{11}=0.22$). 
On the other hand, the bottom Yukawa has a significant contribution of
the pseudoscalar coupling, $b_d$. The effect of this should be visible in the
$\Phi\rightarrow b\bar b$ and $\Phi\rightarrow \tau \bar\tau$ decays.  
With a more relaxed
consideration of the spin and parity measurement of the resonance, we take the other two 
values of $O_{31}=0.32$ and $O_{31}=0.50$, corresponding to the pseudoscalar component of about 
10\% and 25\%, respectively. Here, for small $\tan\beta$ the structure of the top quark 
Yuwaka couplings is modified, with large contribution from $b_u$, while for large
$\tan\beta$, this parameter remains very small. In the former case, the CP-violating
effect is significant in both the production as well as the decay. 

Thus study using
suitably chosen observables at the production level and decay level will be able to
distinguish Model I from Model II, and possibly provide more information about the mixing 
in case of Model II.

\begin{table}
\begin{tabular}{|c|c|c|c|c|c|r|r|r|r|}
\hline
Point&$\tan\beta$
&${\cal O}_{11}$
&${\cal O}_{21}$
&${\cal O}_{31}$
&$Z,~W$&\multicolumn{2}{c|}{top}&\multicolumn{2}{c|}{$b~/~\tau$}\\ \cline{6-10}
&&&&&$c$&$a_{u}$&$b_{u}$&$a_{d}$&$b_{d}$\\ \cline{1-10}
P1&2&0.22& 0.97&0.10 &  0.97 &   1.08 &  $-$0.05 & 0.50  & $-$0.20 \\\cline{1-10}
P2&2&0.22& 0.92&0.32 &  0.92&    1.03&   $-$0.16&  0.50&   $-$0.64\\\cline{1-10}
P3&2&0.22& 0.84&0.50&  0.85&    0.94&   $-$0.25&    0.50 &  $-$1.00\\\cline{1-10} 
P4&20&0.22& 0.97 &0.10&  0.98 &   0.97 &  $-$0.01 & 4.48 &  $-$2.00 \\\cline{1-10}
P5&20&0.22& 0.96 &0.32&  0.93 &   0.92 &  $-$0.02 & 4.48 &  $-$6.40 \\\cline{1-10}
P6&20&0.22& 0.84 &0.50&  0.85 &  0.84 &  $-$0.03 & 4.48 & $-$10.00\\\cline{1-10}
\end{tabular}
\caption{Couplings of the Higgs Bosons ($H_1$) in the CP-violating 2HDM, as defined in
Eq.\,(\ref{eq:cpvmssm_coupl}), for different mixings given by $H_1={\cal O}_{11}~\phi_1+{\cal
O}_{21}~\phi_2+{\cal O}_{31}~A$.
}.
\label{tab:2HDM_coupl}
\end{table}

\section{Process}\label{process}

The process we consider is the associated production of Higgs Boson with $t\bar{t}$ pair in  
$e^{+}e^-$ collision.
As per the Feynman diagrams in Fig.~\ref{fig:0}, the process 
$e^{+}e^- \rightarrow t\bar{t}\Phi$  at leading order proceeds 
(i) through the $s$-channel production of $t\bar{t}$ pair and subsequent raidaton of Higgs 
Boson off either the top quark or the top antiquark, and 
(ii) through Higgsstrahlung process $e^+ e^{-} \rightarrow Z^* \Phi$ followed by 
$t\bar{t}$ pair production from $Z^*$. However, the contribution of the latter channel is confined to few percent\cite{DDGMR} for $\sqrt{s} \leq 1$ TeV.

In our work, we consider an
effective model deviating from the SM through modifications of $f\bar f\Phi$ and
$VV\Phi$ couplings, where $f=t,~b,~\tau$ and $V=Z,~W$.
These changes can be implemented in the {\sc WHIZARD} programme, by suitable
modifications of the programme files of the SM case\footnote{We refer to 
http://feynrules.irmp.ucl.ac.be/wiki/StandardModel in this regard.} 
We have cross checked the correctness of our implementation by verifying the 
results of Ref. \cite{DDGMR} for the process being scrutinized.

It may be noted that the polarization of the top quarks produced here depend on the structure of the Yukawa
coupling. Indeed, spin analysis of the top quark and/or the top antiquark produced in the
process considered is a very useful tool to probe the Yukawa couplings, and thus the CP
nature of the Higgs Boson. One approach is to reconstruct the top
polarization, and study the polarization asymmetry. Such a study was carried out in 
Ref.~\cite{DDGMR}, the result of which we reconfirm. One way to derive the spin
information of the top quark is a study of the distributions of its decay products. The
top quark decays 
nearly 100\% through the electroweak interaction into a $W$ Boson and $b$ quark. 
The decay distribution of the $W$ will provide the information about the top quark
polarization, and hence information about the top quark Yuwaka interaction. We assume that
the gauge structure is the same as that of the SM, and therefore, the $Zt
\bar t$ coupling remains the same as that in the SM. 
In our signal process in Eq. (\ref{signalprocess})
restricting the $Wb$ invariant mass to be that of top quark mass 
guarantees that backgrounds are reduced to a satisfactory level, as will be demonstrated in Section~\ref{results}.

In order to probe the other Yukawa couplings, we analyse the decay of the Higgs Boson.
The main decay modes of the Higgs of mass around 125\,GeV are $b\bar{b}$, $W W^{*}$ and 
$\tau^{+}\tau^{-}$
with branching fractions of 57.7\%, 21.5\% and 6.32\% respectively. Since $W$'s couple
only to the CP-even component of the Higgs Boson, the $WW$ channel can give only very 
limited information about CP mixing. Again, the branching fraction into $\tau\bar\tau$ 
pair is very small, and therefore, we will mainly discuss the $b\bar b$ decay channel.
In order to reduce the complexity, when $\Phi$ decay is considered, we assume that the top
quark and top antiquark are fully reconstructed.
This leaves us with a final state of four particles in our signal process, 
 $e^+e^-\rightarrow t \bar t \Phi\rightarrow t \bar{t} b\bar{b}$. 
Here again, we will restrict the invariant mass of $b\bar b$ to be around 125\,GeV to
eliminate the background.

\section{Results}\label{results}

We first consider Model I. As discussed in Section~\ref{formalism}, there is only one
independent parameter, which is taken to be $b$, which can vary from $0$ to $1$. 
$b=0$ corresponds to the purely CP-even Higgs Bosons, like that in the SM, while $b=1$ 
corresponds to the purely CP-odd Higgs Boson.

\begin{figure}[!t]\centering
\vspace*{-1cm}
\begin{tabular}{c c c}
\includegraphics[angle=0,width=80mm]{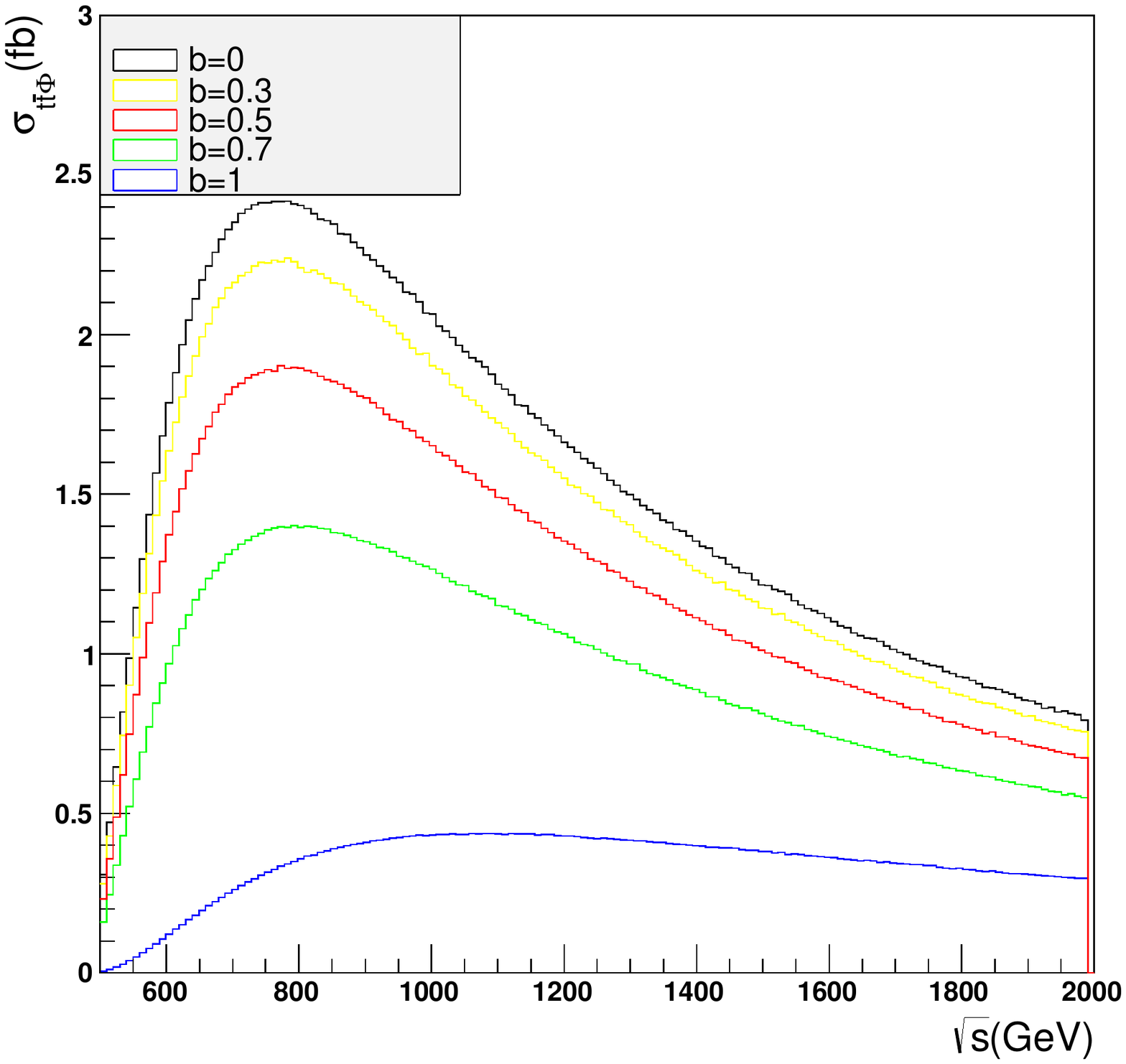} &
\hspace*{-25mm}
\includegraphics[angle=0,width=80mm]{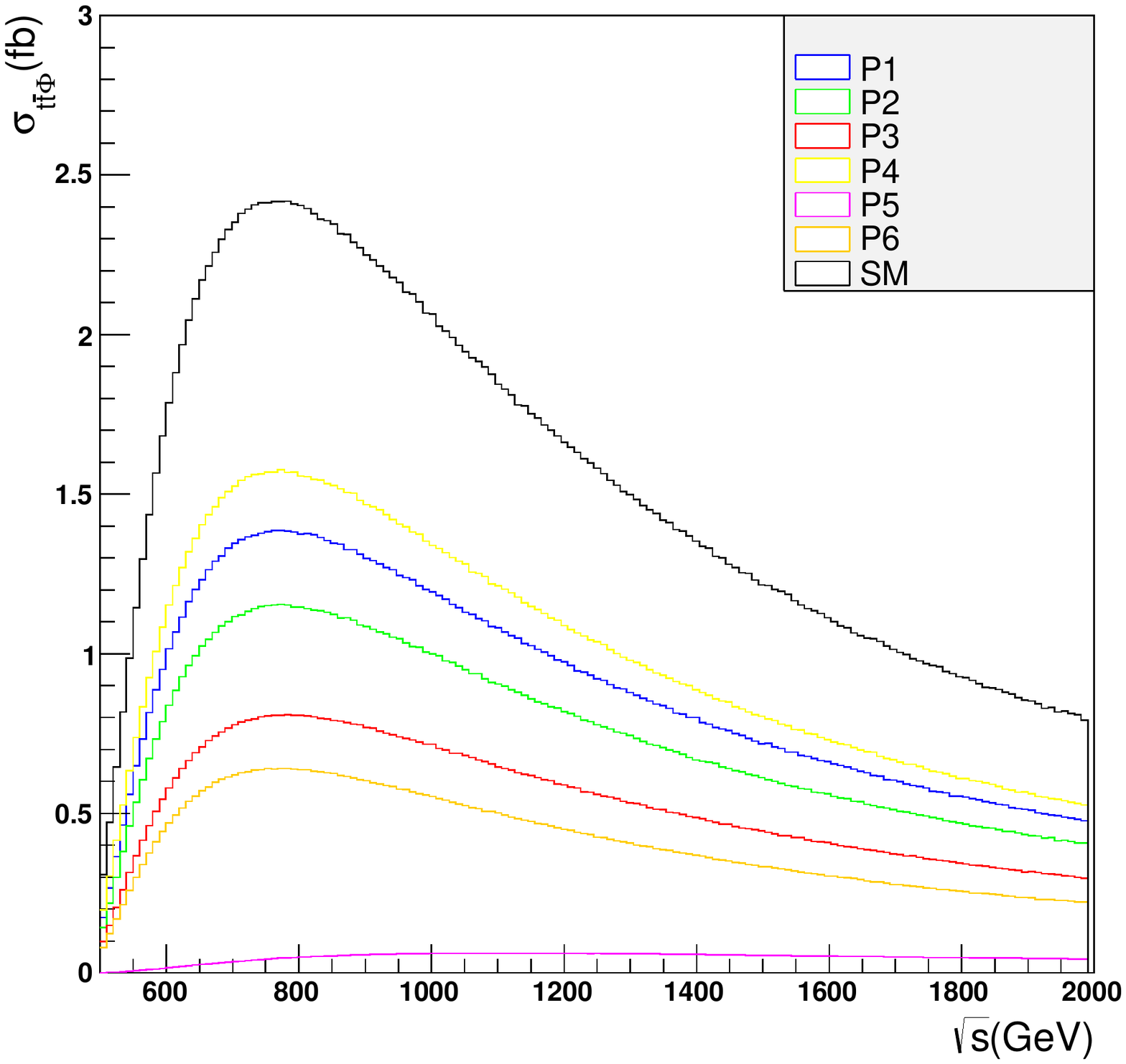} &
\hspace*{-25mm}
\includegraphics[angle=0,width=80mm]{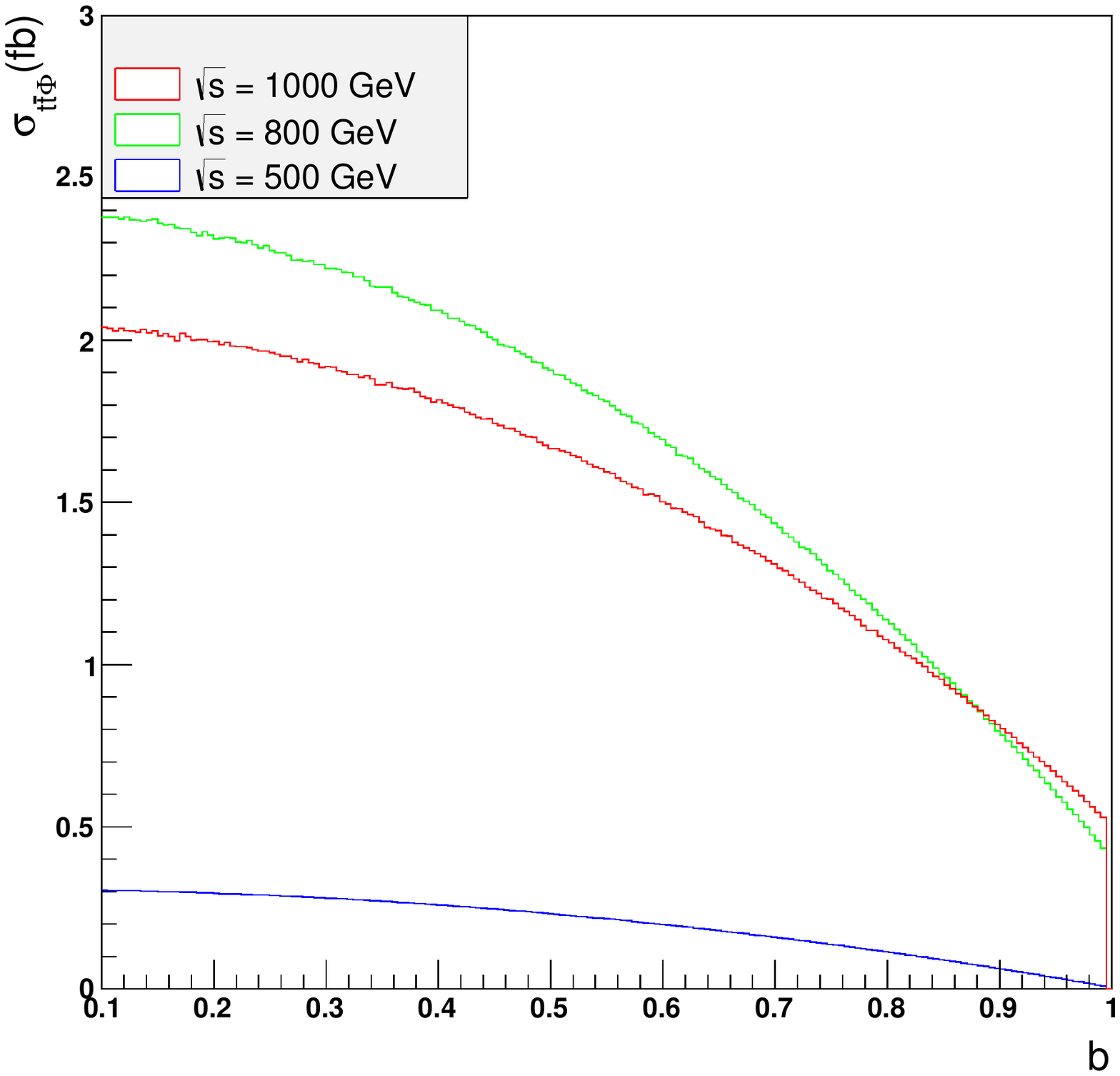} 
\end{tabular}
\vskip 1mm
\caption{ $\sqrt{s}$ vs.  $t\bar{t}\Phi$ production cross section in Model I(left fig.), in Model II(middle fig.), and b vs. $t\bar{t}\Phi$ production 
cross section in Model I (right fig.) for various values of $\sqrt{s}$.
}
\label{fig:cs_roots}
\end{figure}

The CP nature of the Higgs Boson affects  the total production cross section, which is 
the first observable that we consider here. We plot this in Fig.~\ref{fig:cs_roots}
(left) for some values of $b$. The mass of the Higgs Boson is taken to be 125\,GeV in 
all our analyses. As we can see the total cross section is indeed sensitive to the
parameter $b$, with maximum deviations from SM for non-zero $b$ values at centre of
mass of 800\,GeV. Notice that the deviations are not linear in $b$, with about 8\%
deviation for $b=0.3$, which goes upto about 21\% for $b=0.5$. In
Fig.~\ref{fig:cs_roots} (right) we consider the cross section at $\sqrt{s}=
500,~~800,~~1000$\,GeV plotted against the parameter $b$.

\begin{figure}[!t]\centering
\begin{tabular}{c c} 
\includegraphics[angle=0,width=100mm]{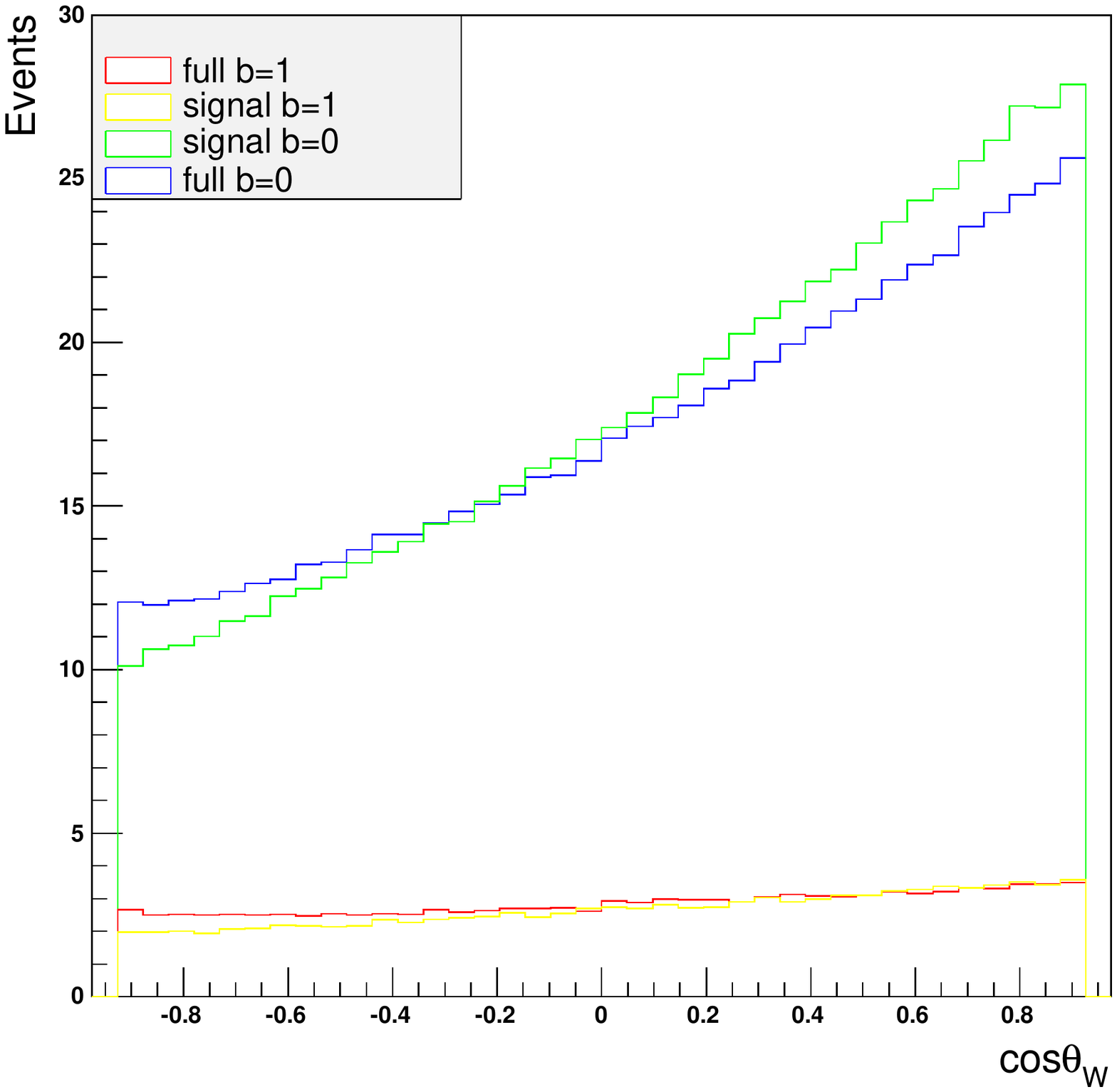} &
\includegraphics[angle=0,width=100mm]{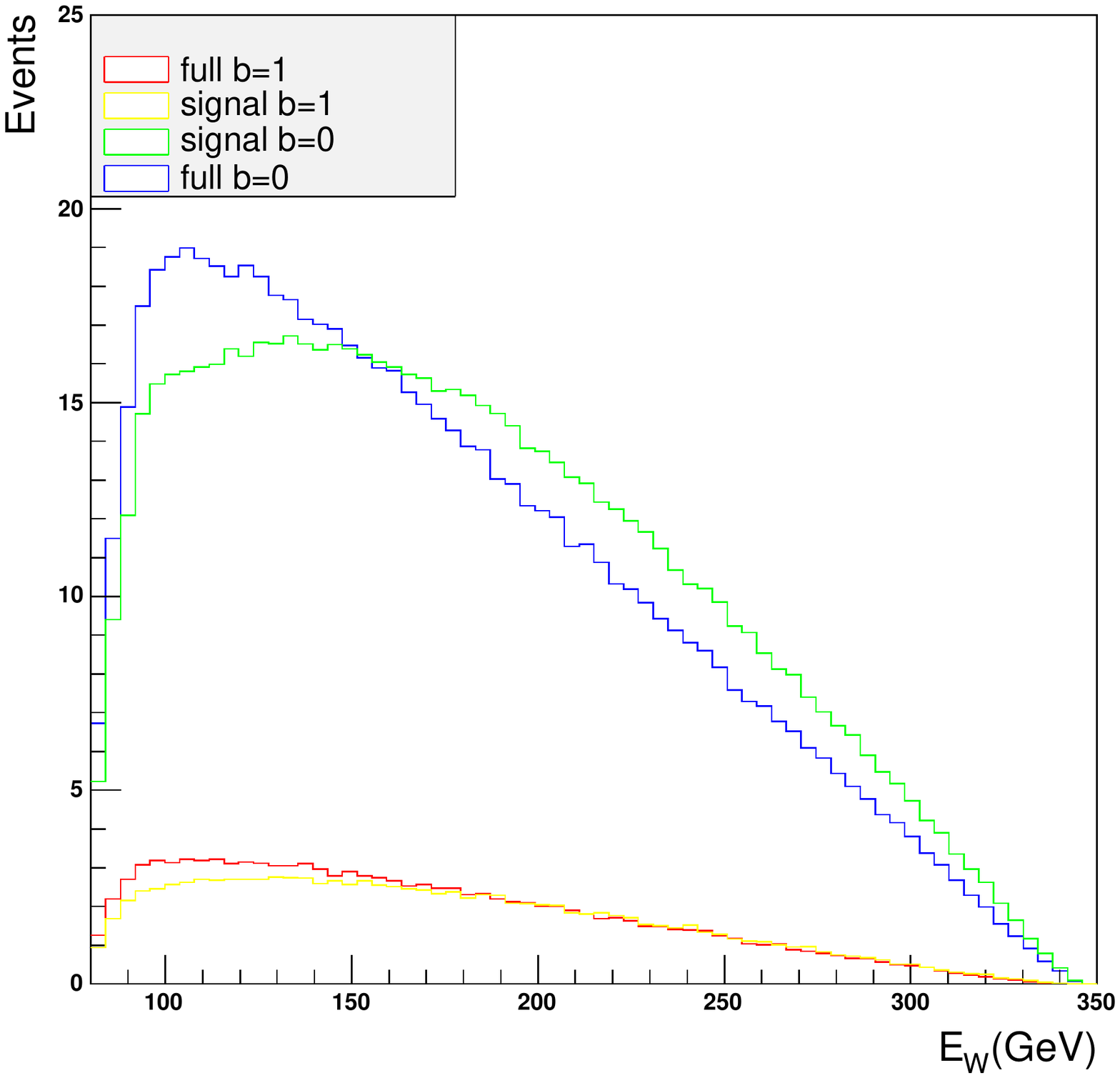}\\
\end{tabular}
\caption{Model I: Polar angle(left fig.) and energy distribution(right fig.) of  $W^+$ at $\sqrt{s}=800$ GeV with an integrated
luminosity of $300~{\rm fb}^{-1}$ in the case of full and signal process for extreme $b$ values. No kinematical cuts have been imposed here.}
\label{fig.3}
\end{figure}

\begin{figure}[!t]\centering
\begin{tabular}{c c} 
\includegraphics[angle=0,width=100mm]{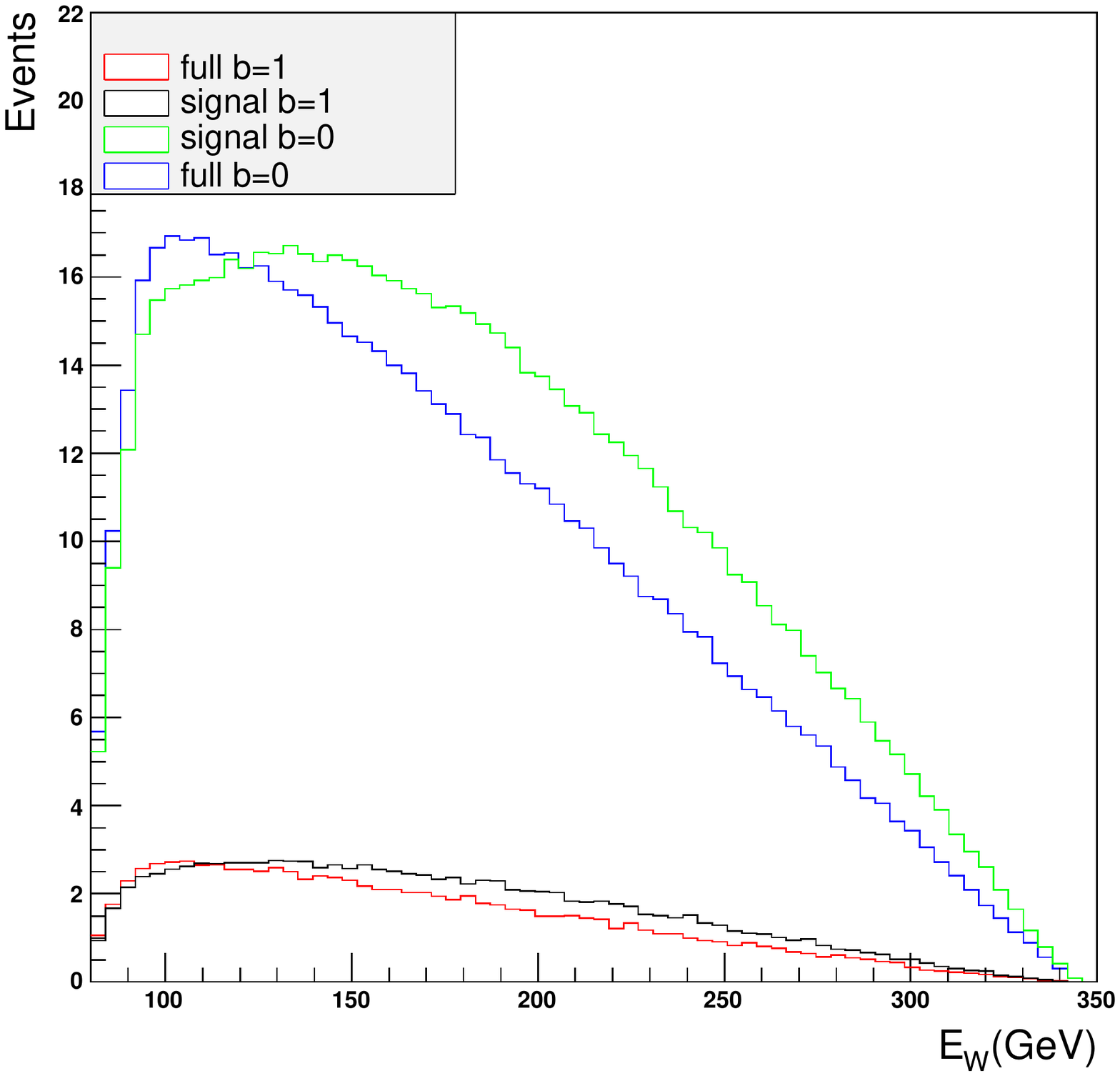} &
\includegraphics[angle=0,width=100mm]{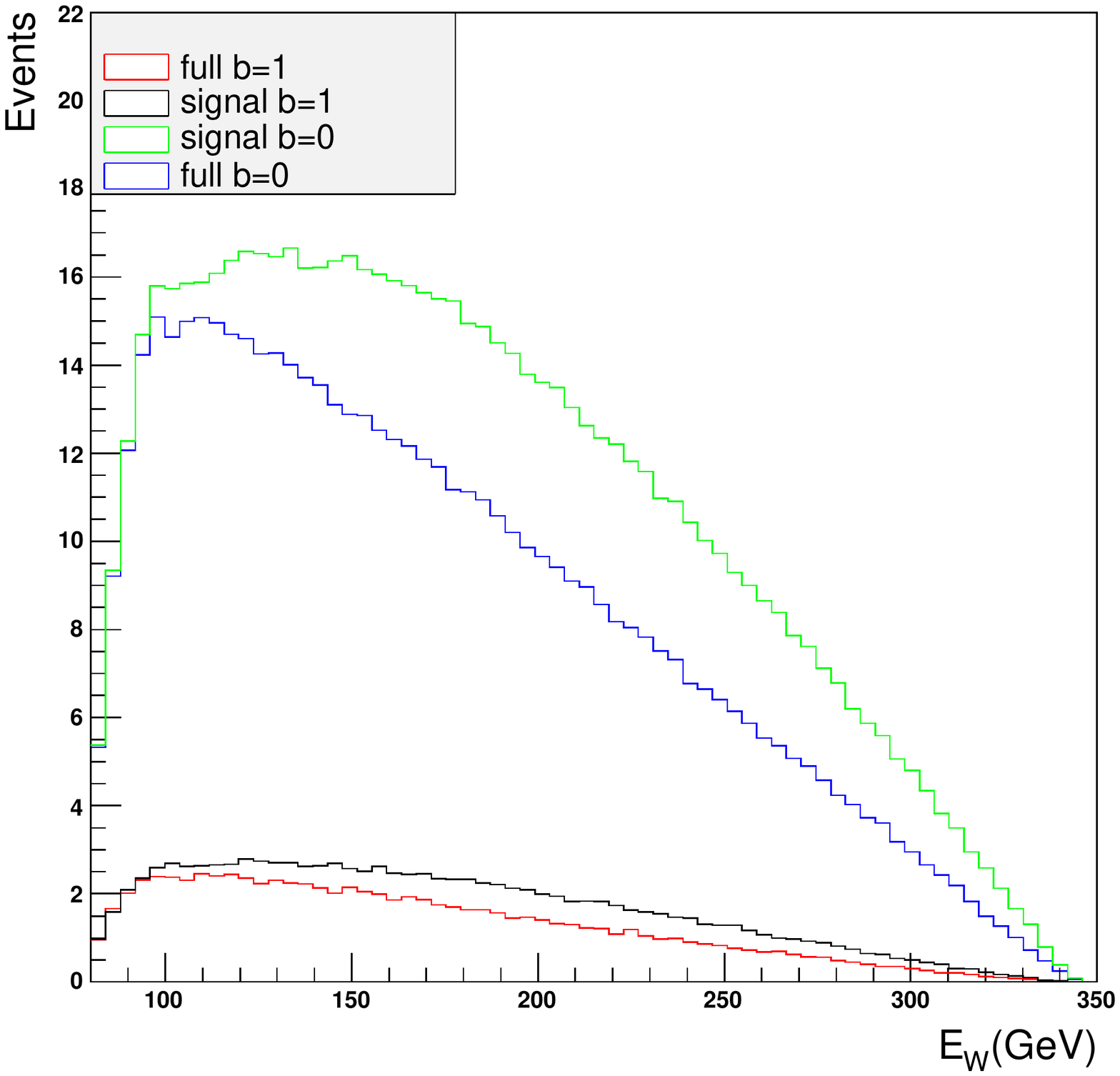}
\end{tabular}
\caption{Model I: Energy distribution with a polar angle cut(left fig.) and $M_{bW}$ 
cut(right fig.) of  $W^+$ at $\sqrt{s}=800$\,GeV with an integrated luminosity of 
$300~{\rm fb}^{-1}$ in the case of full process for extreme $b$ values. }
\label{fig.4}
\end{figure}

The top quark polarization in the process studied is decided by the 
CP properties of the Higgs Boson produced. A top polarization asymmetry measurement, as presented in
\cite{DDGMR} clearly shows the advantage of this observable in identifying the CP
properties of the Higgs Boson. 
In the present work, we go beyond the production process to study the decay products of 
the top quark to analyse the $tt\Phi$ coupling, and thus the CP properties of the Higgs 
Boson. It is important for us to point out the differences between the present work and 
that of Ref. \cite{GHMRS}.  While we have considered in detail the decay distributions 
in contrast to that work, we do not attempt to find, for example, the reach of the 
parameters $a$ and $b$ at the ILC, but rather we aim to illustrate the sensitivity of ILC 
within a realisitic model, assuming specific values of the parameters considered in the 
light of the LHC discovery.  
It is well known that the decay distributions can be used as the top spin analysers. 
As the cross section peaks at around $\sqrt{s}=800$\,GeV, gaining in statistics, 
we will consider the distributions at this centre of mass energy.
While considering the final state, $WWb\bar b$ in this case, we need to worry 
about the background. In the following we call the signal along with the the
background as the full process. 
In Fig.~\ref{fig.3} we plot the angular and energy distributions of the $W^+$ of the signal process 
as well as the full process for the two extreme cases of purely scalar Higgs Boson ($b=0$), like the SM Higgs
Boson, and the purely psudoscalar Higgs Boson ($b=1$). As mentioned in Section~\ref{formalism}, it is very likely
that the value of $b$ is close to zero. In such case the curves will stay closer to the SM curve. 
The angular distribution clearly shows a different forward-backward asymmetry in the case
of signal process in comparison to the background process. The signal process is more peaked in
the forward direction, indicating that an appropriate cut on
$\cos\theta$ can increase the signal over the background.  The distribution becomes 
more flat with increasing pseudo-scalar composition. A forward-backward asymmetry 
to be described later will use this feature to extract information on the reach of $b$ at ILC
through the process considered. 
The energy distribution of the $W^+$ is also highly sensitive to
the parameter $b$. The larger deviation in the central region indicates that a
suitable kinematic cuts can enhance the sensitivity further. 

It is clear from Fig.\ref{fig.3} that the background is very small, and even
this small background can be controlled by an angular cut. Since in the signal
process $W^+b$ is a product of the top quark decay, the invariant mass of this pair
is expected to lie around top quark mass. Thus, a cut on this variable is an
efficient way of controlling or even eliminating the background. 
We apply a cut of $170$\,GeV  $< M_{bW} < $  $175$\,GeV to suppress the backgrounds in the full
process. The resultant energy distributions of the $W^+$  with angular and $M_{bW}$ cut 
are given in Fig.~\ref{fig.4}.  Thus,
in the rest of the discussion  we use only the signal process.

Taking cue from the angular distributions, we construct the forward-backward asymmetry
and present this in Table~\ref{tab:AFB_M1} for different parameter values at the centre 
of mass of 800 GeV. The asymmetry is significant only for large values of $b$ with
about 6\% deviation from the SM case at $b=0.7$ and about 40\% deviation for $b=1$. 
The beam polarization at ILC is expected to play an important role in studying the effects of new physics. This machine
is supposed to provide high degree of polarization in longitudinal and transverse mode. To discuss the effects of 
initial longitudinal beam polarization we have generated  similar distributions as we have for unpolarized case. For our
study we have used a realistic 80\% electron($e^-$)  and 60\% positron($e^+$) beam 
polarization.  Since the final state particles do not have any common interaction
vertex with the intial particles, the beam polarization  is not expected to have any
feature qualitatively different from that of the case of unpolarized beam. At the same
time, in the present case the polarization helps improve the statistics. The
forward-backward asymmetry of the $W$ is not affected by beam polarization, 
as can be see from Table~\ref{tab:AFB_M1}.

\begin{table}
\begin{tabular}{|c|c|c|c|c|}
\hline
$b$&$N_{tot}^{\rm unpol}$&$A_{FB}^{\rm unpol}$&$N_{tot}^{\rm pol}$&$A_{FB}^{\rm pol}$\\ \cline{1-5}
0&$720$&$0.269$&$1440$&$0.213$\\\cline{1-5}
0.3&$665$&$0.264$&$1330$&$0.212$\\\cline{1-5}
0.5&$567$&$0.261$&$1133$&$0.209$\\\cline{1-5}
0.7&$420$&$0.254$&$839$&$0.205$\\\cline{1-5}
1&$107$&$0.166$&$214$&$0.133$\\\cline{1-5}
\hline
\end{tabular}
\caption{Forward-backward asymmetry of the $W^+$ and the total number of
events in unpolarized(2nd-3rd column)
and longitudinal polarized(4th-5th column) case for
an integrated luminosity of 300 fb$^{-1}$ with different values of $b$
corresponding to Model I.
}
\label{tab:AFB_M1}
\end{table}

\begin{table}
\begin{tabular}{|c|c|c|c|c|}
\hline
$Case$&$N_{tot}^{\rm unpol}$&$A_{FB}^{\rm unpol}$&$N_{tot}^{\rm pol}$&$A_{FB}^{\rm pol}$\\ \cline{1-5}
P1 &$836$&$0.253$&$1669$&$0.212$\\\cline{1-5}
P2 &$762$&$0.252$&$1523$&$0.211$\\\cline{1-5}
P3 &$640$&$0.252$&$1278$&$0.210$\\\cline{1-5}
P4 &$679$&$0.253$&$1356$&$0.213$\\\cline{1-5}
P5 &$610$&$0.254$&$1219$&$0.212$\\\cline{1-5}
P6 &$509$&$0.253$&$1017$&$0.211$\\\cline{1-5}
\hline
\end{tabular}
\caption{Forward-backward asymmetry of the $W+$ and the total number of
events in unpolarized(2nd-3rd column)
and longitudinal polarized(4th-5th column) case for
an integrated luminosity of 300 fb$^{-1}$ with different cases
corresponding to Model II are tabulated.
}
\label{tab:AFB_M2}
\end{table}

\begin{figure}[!t]\centering
\begin{tabular}{c c} 
\includegraphics[angle=0,width=100mm]{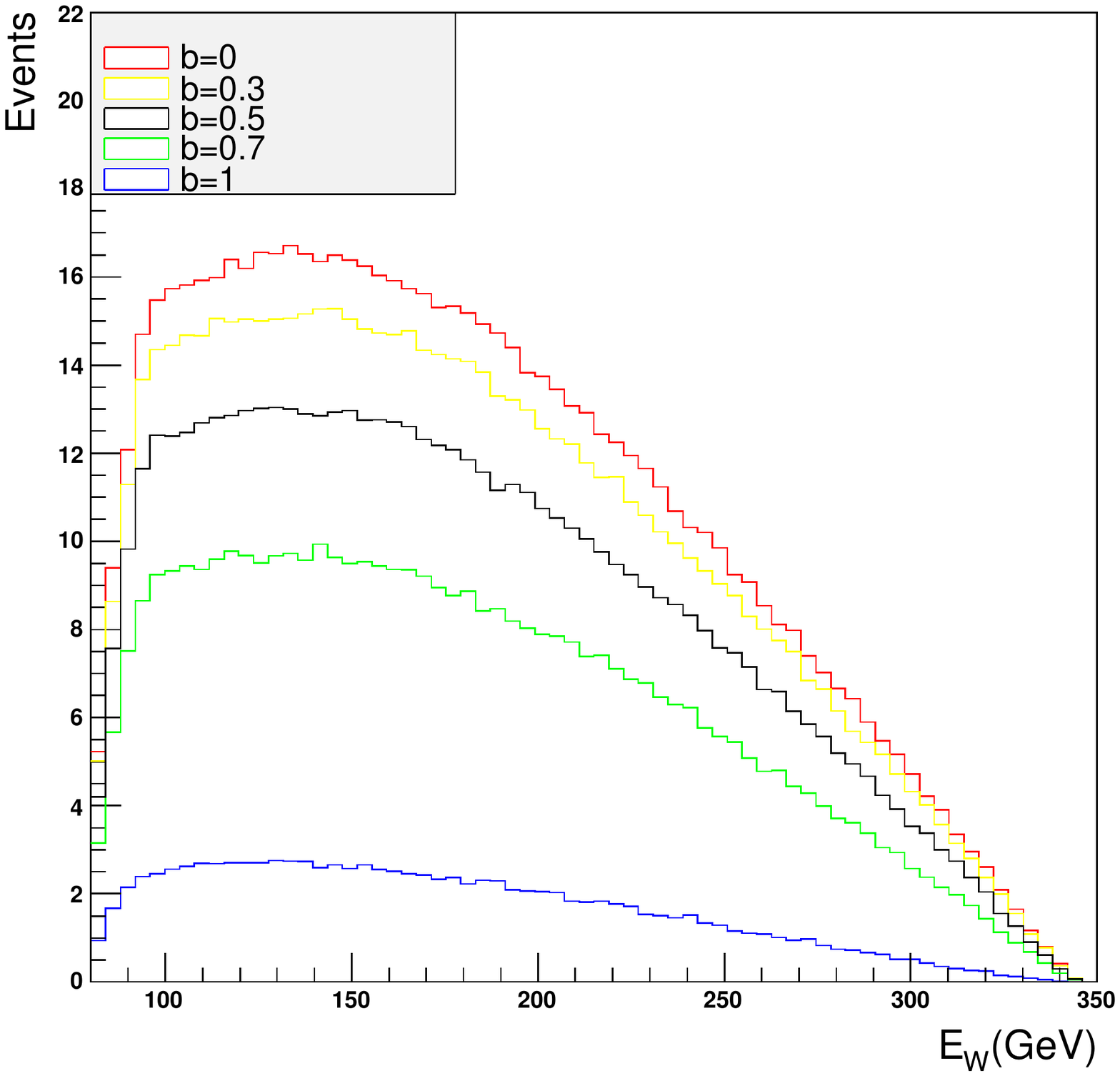} &
\includegraphics[angle=0,width=100mm]{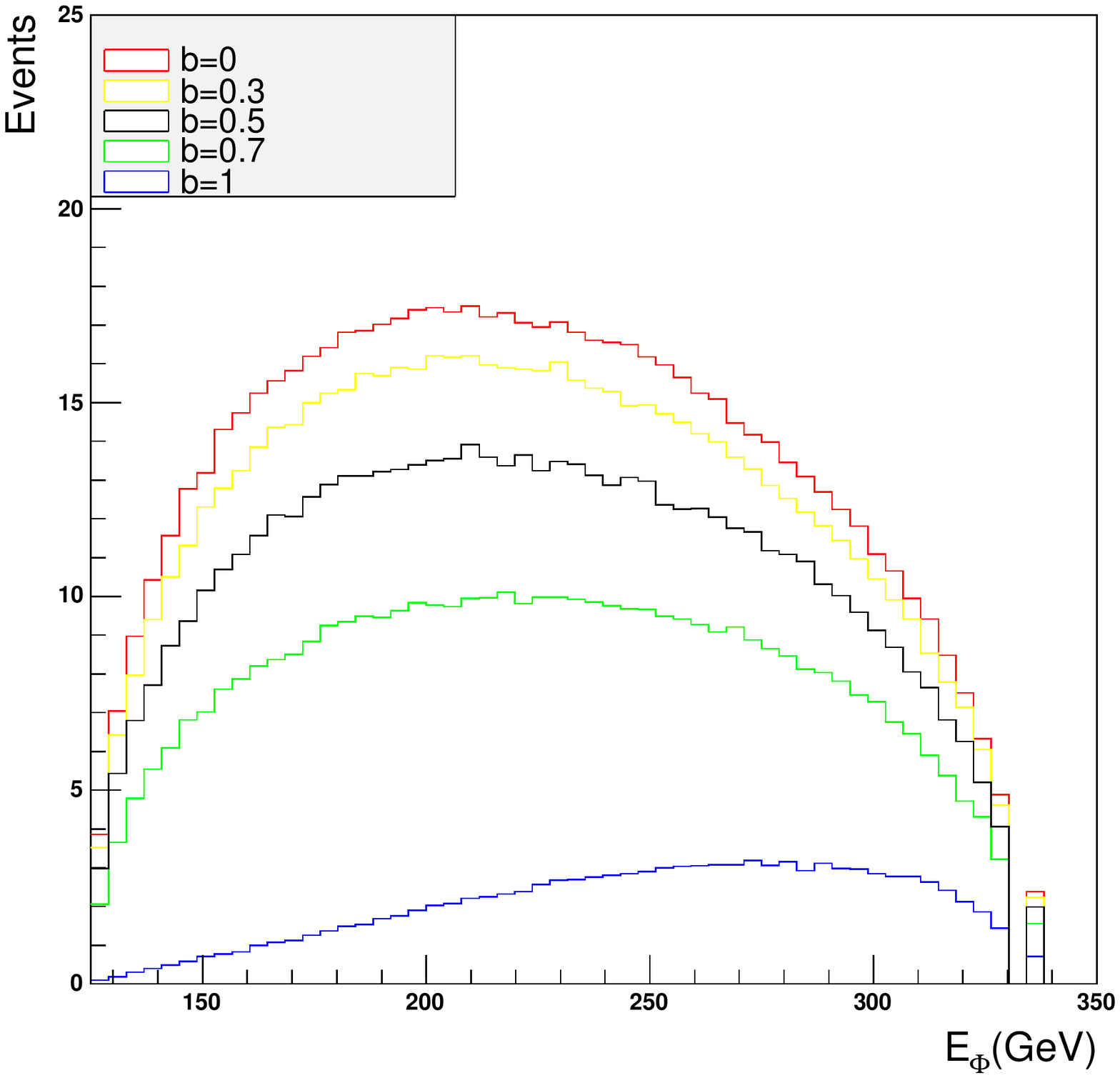}
\end{tabular}
\caption{Model I: Energy distributions of $W^+$ and $\Phi$ in 
$e^+e^-\rightarrow t\bar t \Phi\rightarrow b\bar b W^+W^- \Phi$ with unpolarized beams
at $\sqrt{s}=800$\,GeV with an integrated
luminosity of $300~{\rm fb}^{-1}$ for different values of $b$.}
\label{fig.5}
\end{figure}

\begin{figure}[!t]\centering
\begin{tabular}{c c} 
\includegraphics[angle=0,width=100mm]{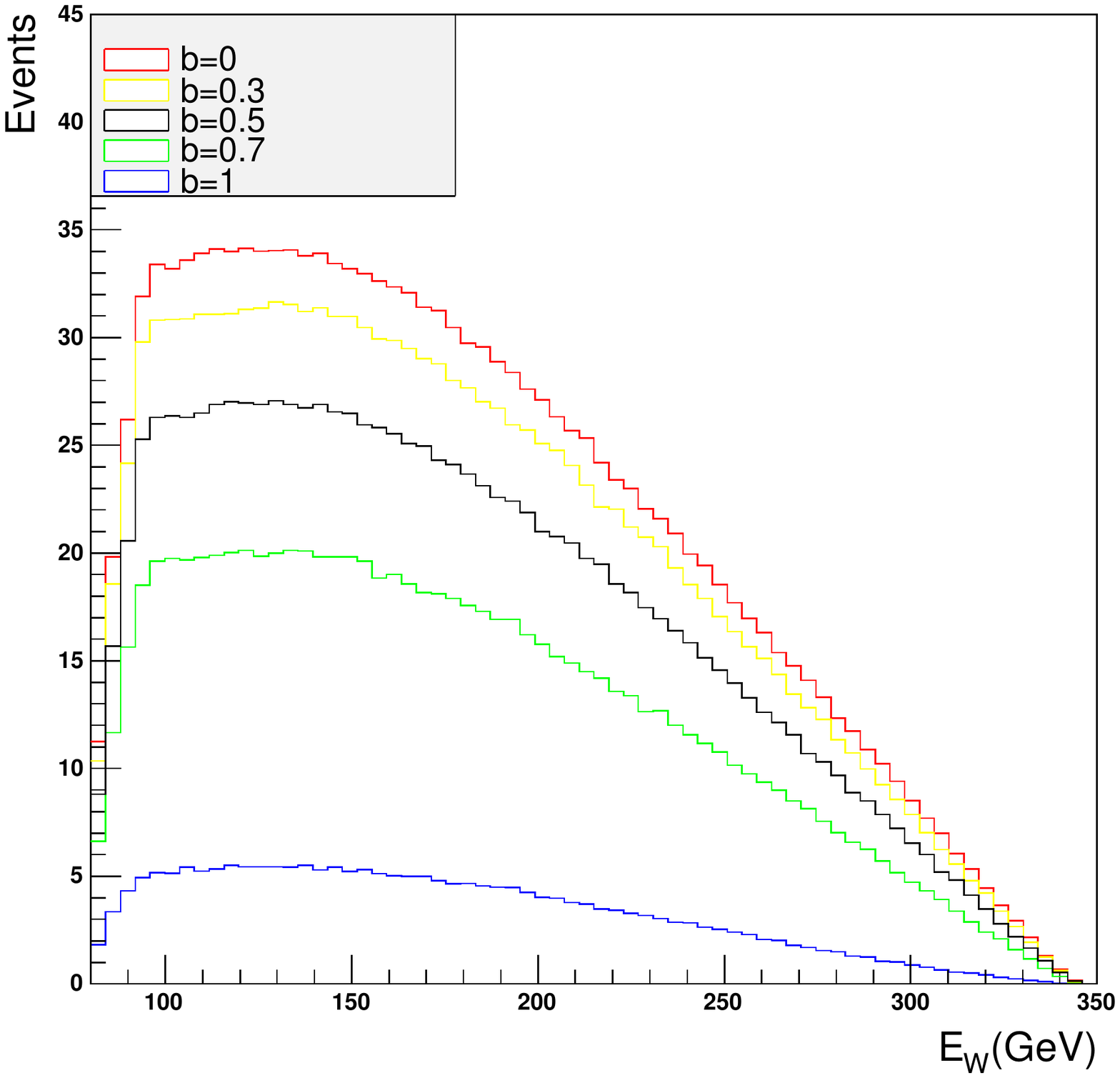} &
\includegraphics[angle=0,width=100mm]{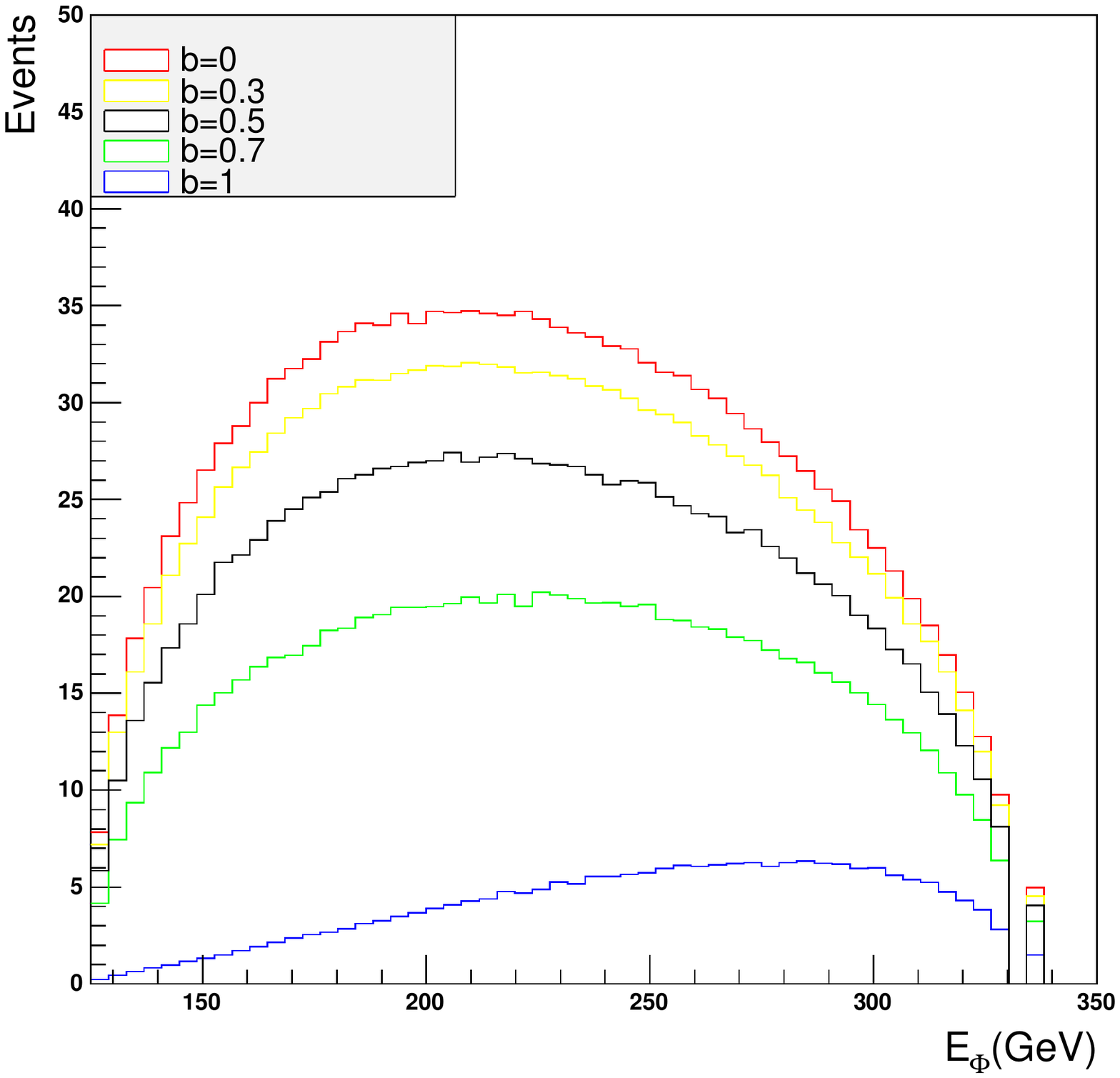}
\end{tabular}
\caption{Model I: Energy distributions of $W^+$ and $\Phi$ in 
$e^+e^-\rightarrow t\bar t \Phi \rightarrow b\bar b W^+W^- \Phi$
at $\sqrt{s}=800$\,GeV with an integrated
luminosity of $300~{\rm fb}^{-1}$ for different values of $b$. Beam polarizations of
80\% electron beam polarization and 60\% positron beam polarization are considered.}
\label{fig.6}
\end{figure}

In Fig.~\ref{fig.5} we present energy distributions of the $W^+$ and Higgs Boson for different values of the parameter $b$.
We note that significant deviation is present only for values of $b=0.5$ and beyond. 
Apart from the reduction in the distribution for the whole range of the energy
values, a shift in the maximum towards higher energy values for larger $b$ values is
noted in the case of energy distribution of the Higgs Boson. This may give an additional 
handle to pinpoint the contribution of the pseudoscalar component in the Higgs Boson.
In Fig.~\ref{fig.6} we consider the energy distribution of the $W^+$ and the Higgs
Boson in the presence of beam polarization. Firstly we note that the total number of 
events is almost doubled compared to the case of unpolarized beams. 
As a consequence, the sensitivity is
improved, and it is possible to have more than $1\sigma$ deviation (considering only
statistical uncertainty) for smaller value of $b$ compared to the case of
unpolarized beams. While the reach of $b$ is certainly improved with beam
polarization, we see that it is still not really possible to probe pseudoscalar 
admixtures of a few percent or even up to 10-20\%.

\begin{figure}[!t]\centering
\begin{tabular}{c c c} 
\includegraphics[angle=0,width=80mm]{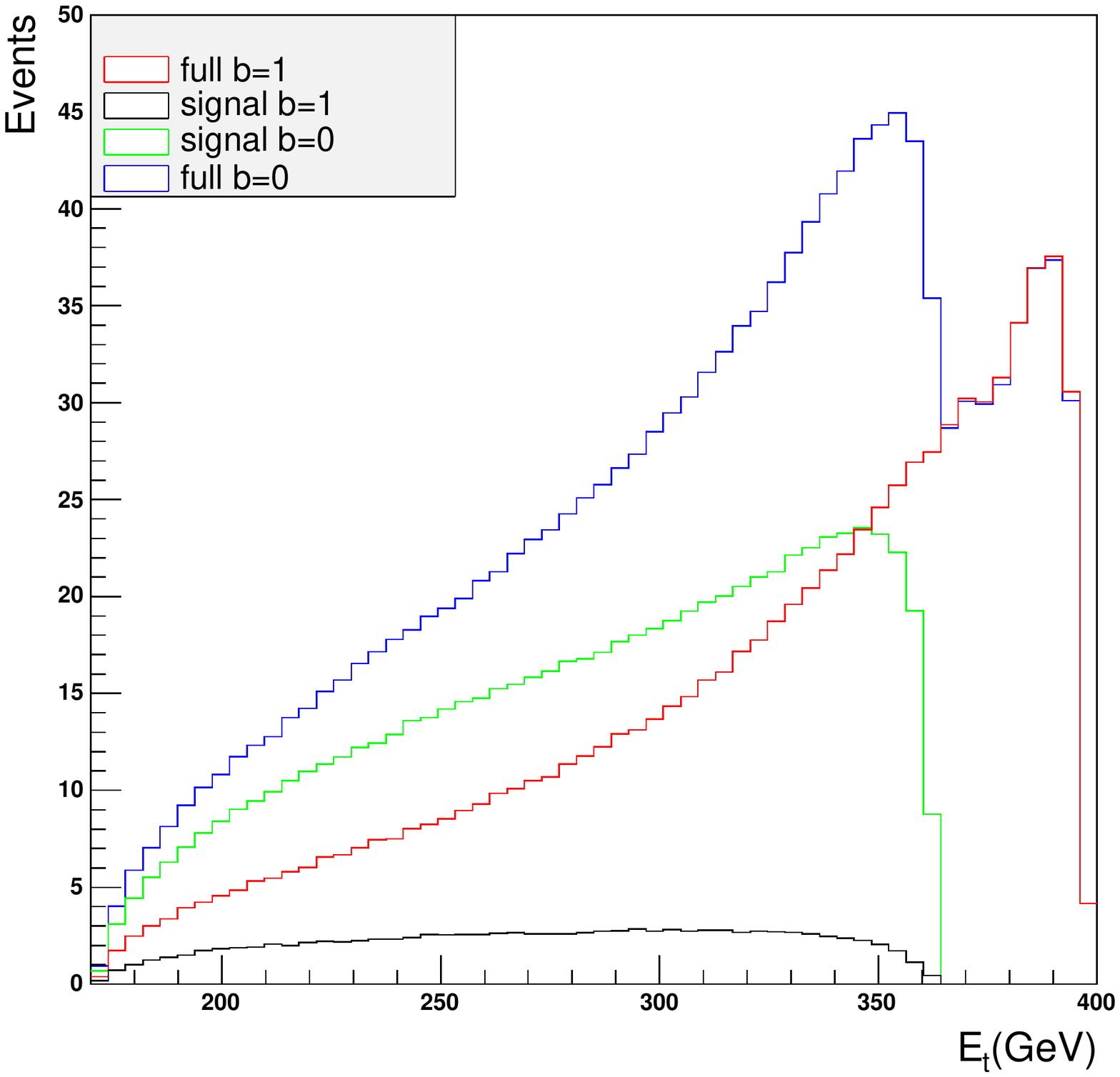} &
\hspace*{-25mm}
\includegraphics[angle=0,width=80mm]{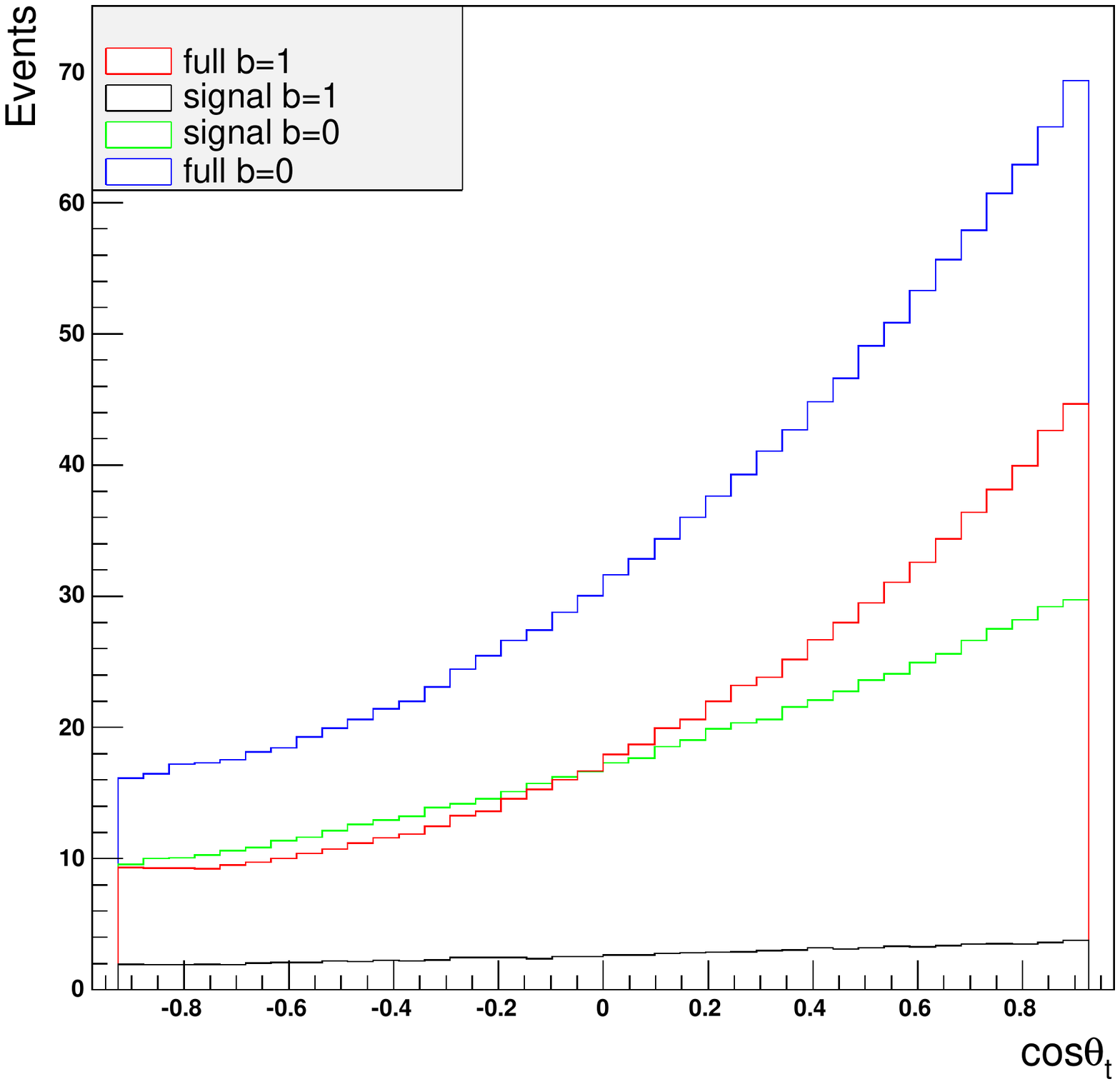}&
\hspace*{-25mm}
\includegraphics[angle=0,width=80mm]{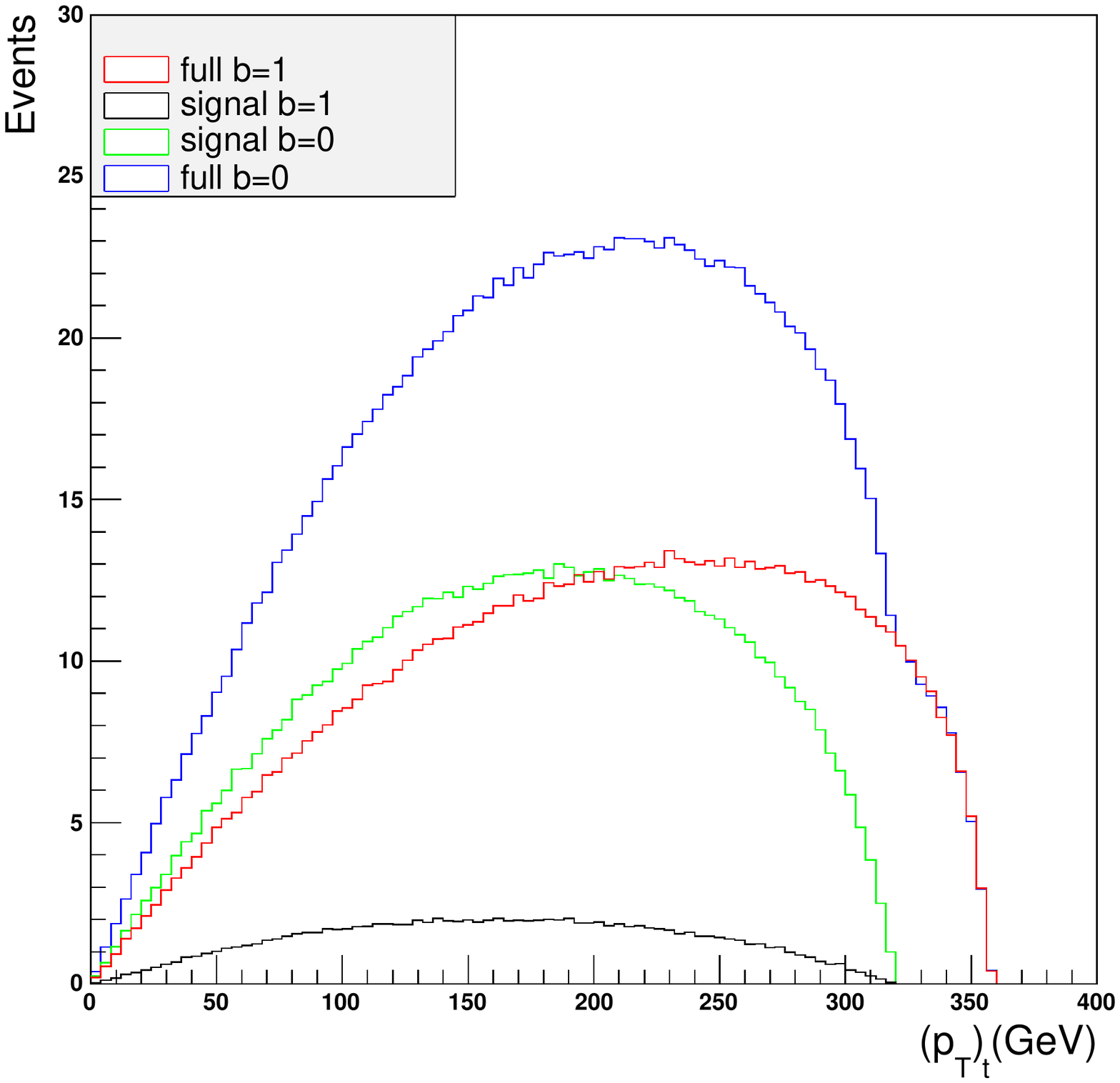}\\
\includegraphics[angle=0,width=85mm]{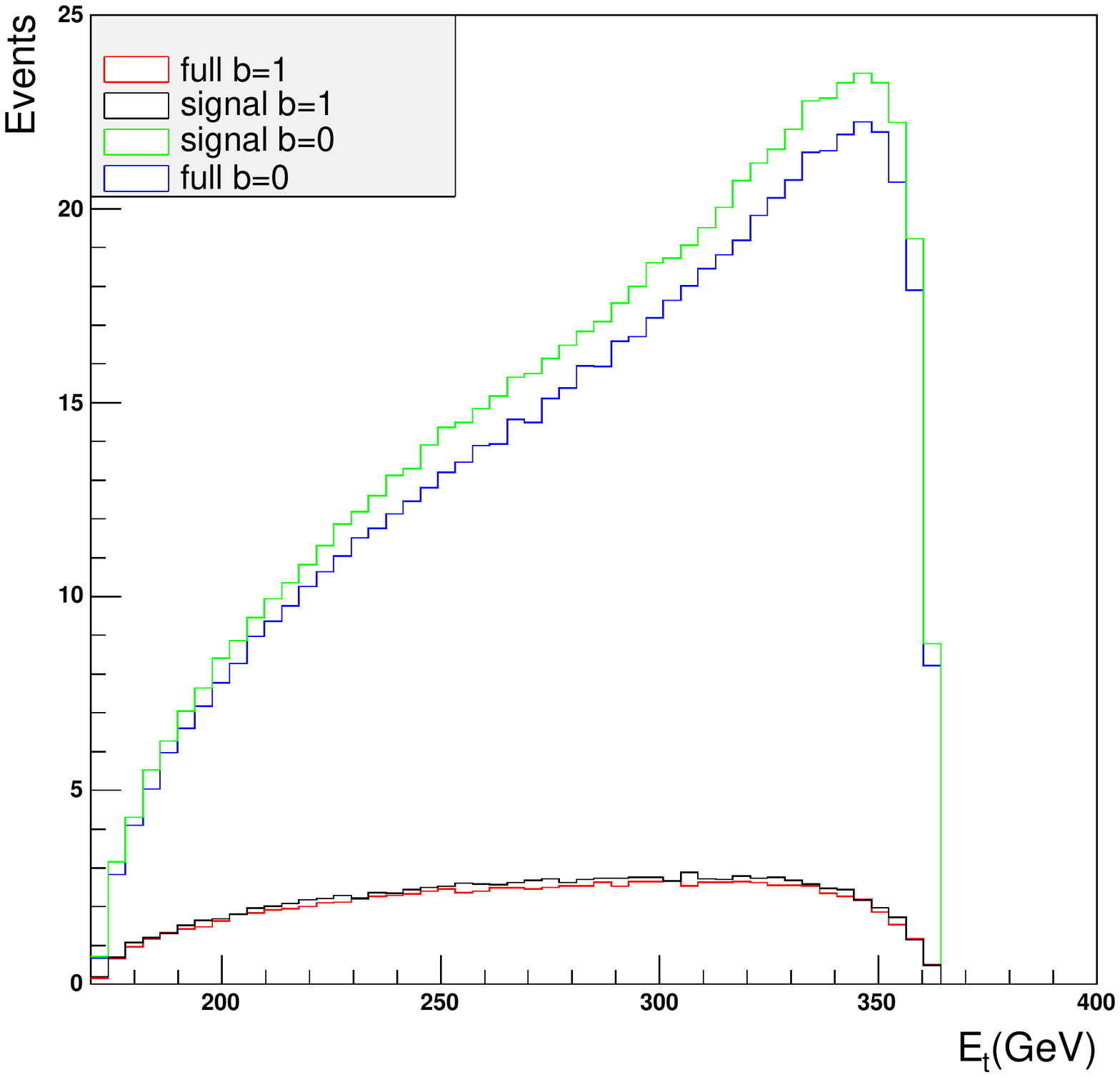} &
\hspace*{-25mm}
\includegraphics[angle=0,width=85mm]{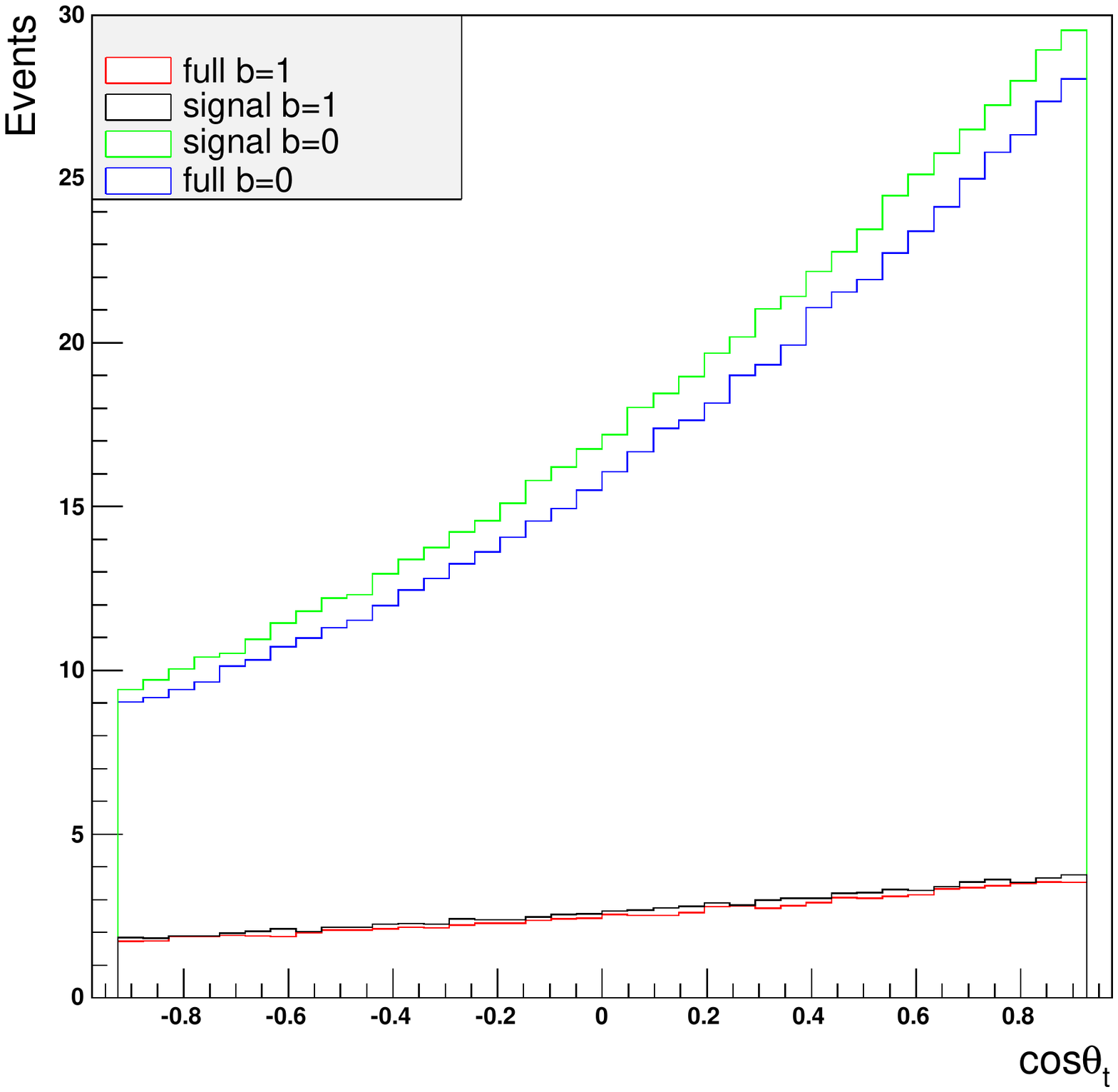}&
\hspace*{-25mm}
\includegraphics[angle=0,width=85mm]{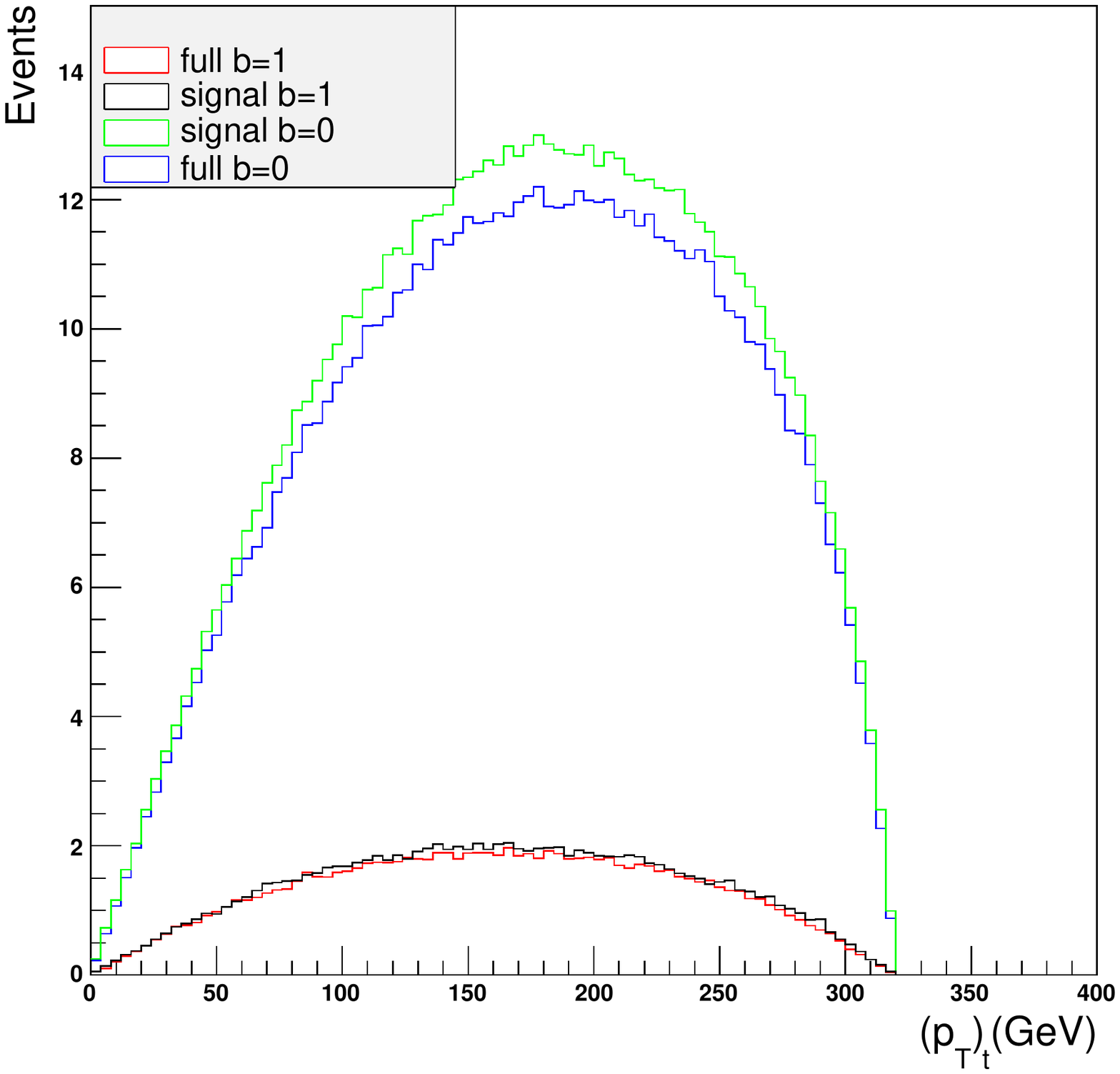}\\
\end{tabular}
\caption{Model I: Energy(left fig.), Polar angle(middle fig.) and $pT$ distribution(right fig.) of top quark
at $\sqrt{s}=800$\,GeV with an integrated
luminosity of $300~{\rm fb}^{-1}$ for full and signal process using extrerme $b$ values.
The top row is without any kinematic cuts, while in the case of the second row
 a cut of $ 124$\,GeV $\leq M_{b\bar{b}} \leq 126$\,GeV have been imposed.}
\label{fig.7}
\end{figure}

\begin{figure}[!t]\centering
\begin{tabular}{c c c}  
\includegraphics[angle=0,width=80mm]{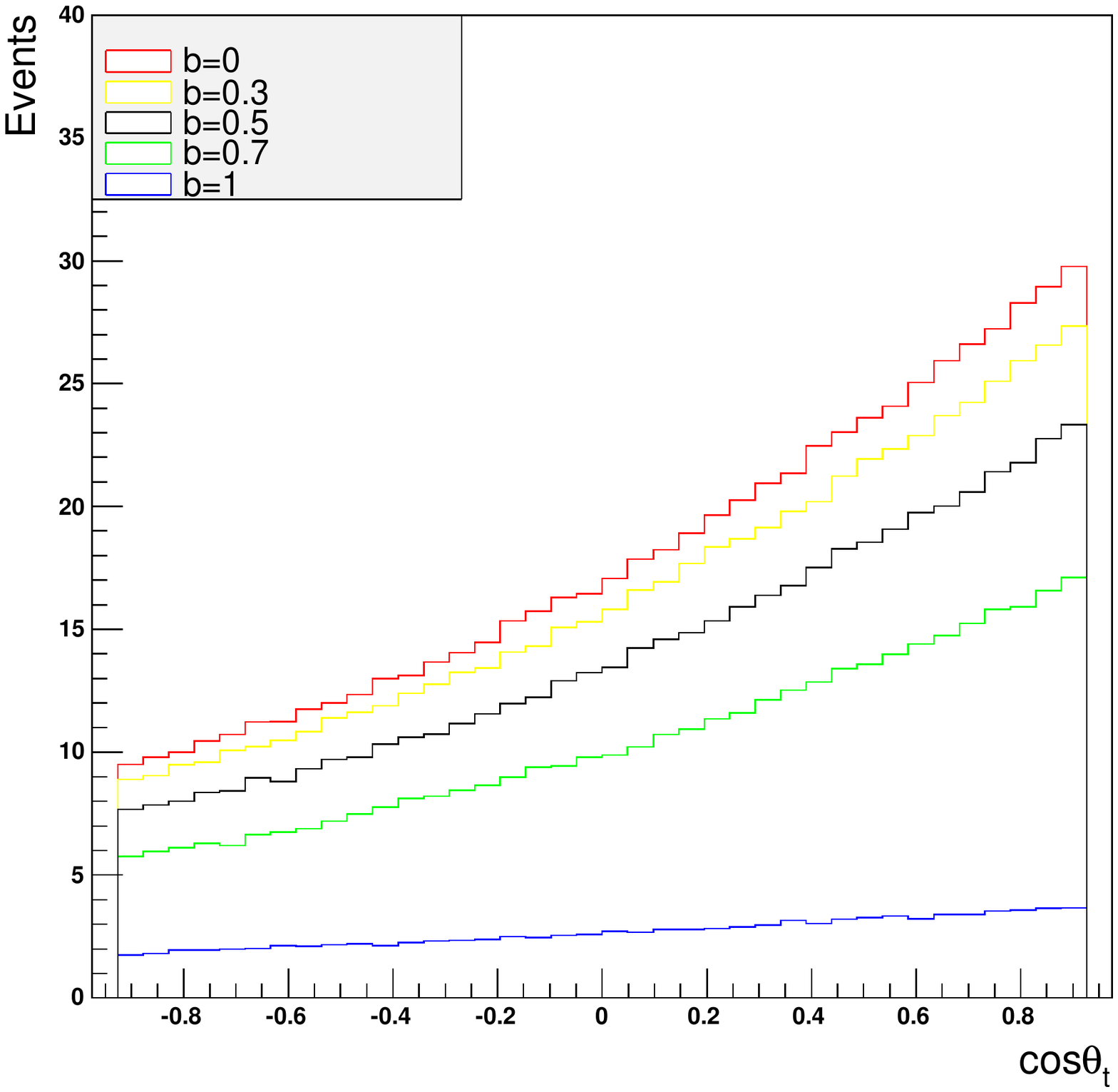} &
\hspace*{-25mm}
\includegraphics[angle=0,width=80mm]{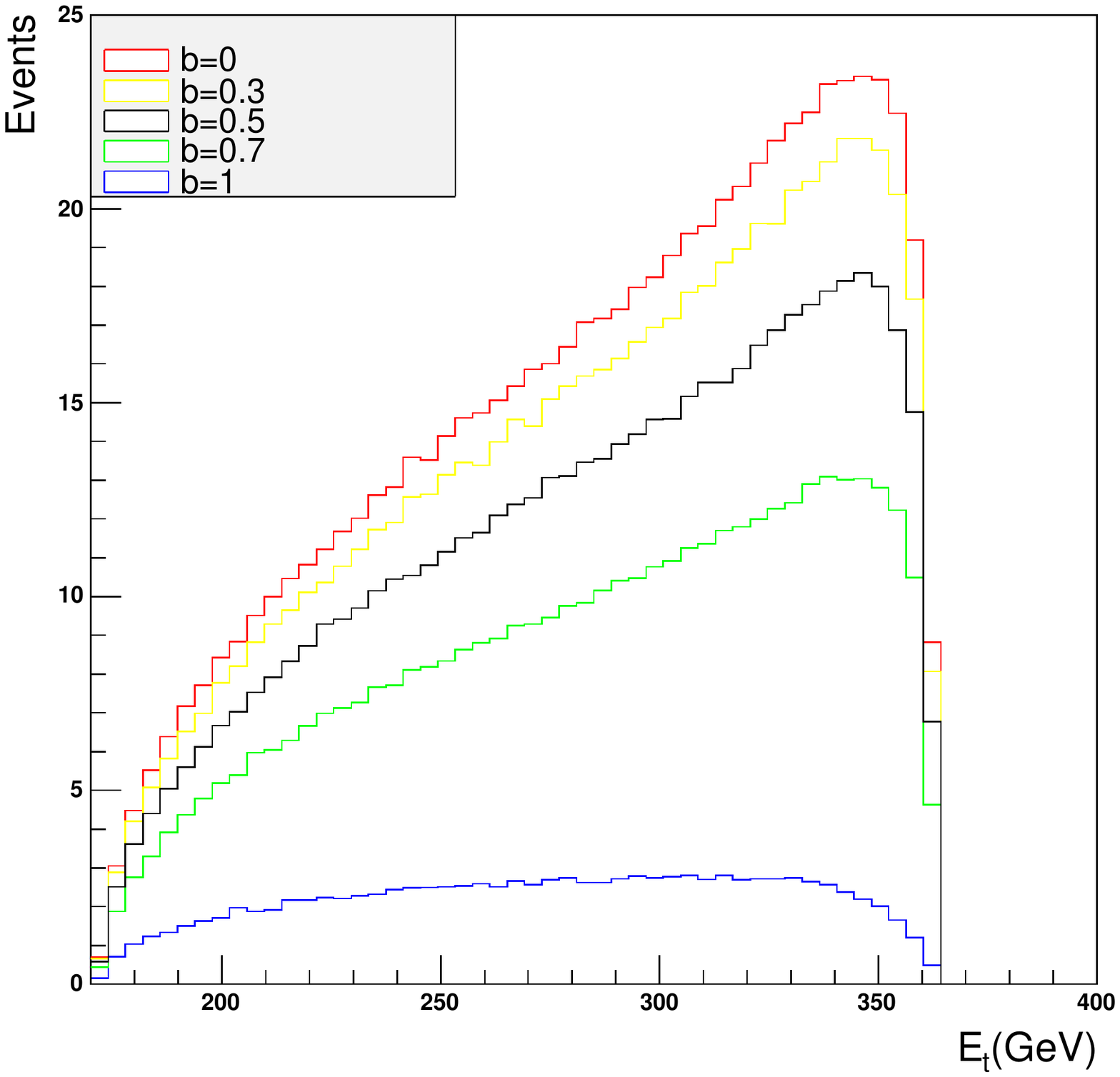} &
\hspace*{-25mm}
\includegraphics[angle=0,width=80mm]{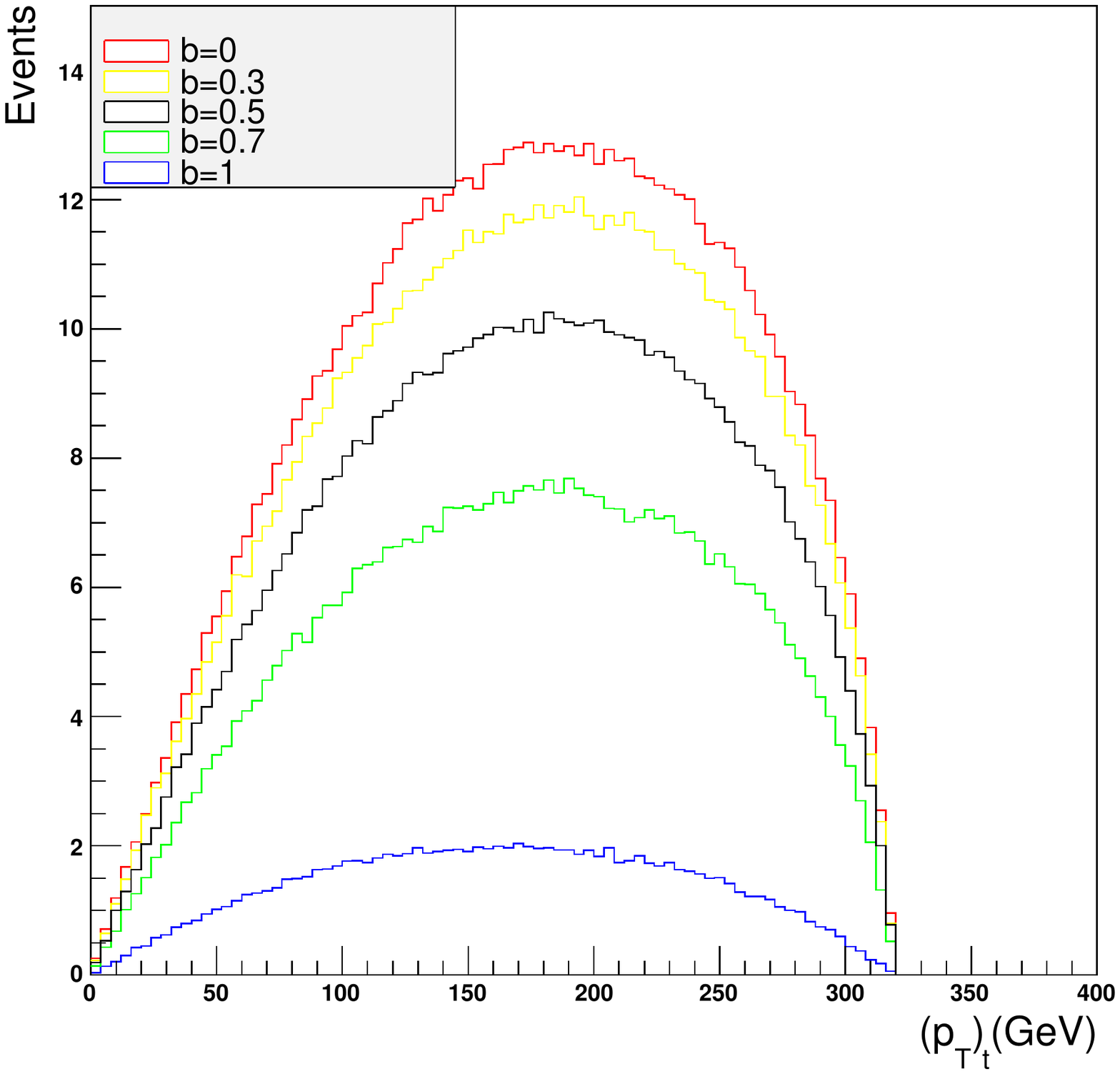} \\
\includegraphics[angle=0,width=80mm]{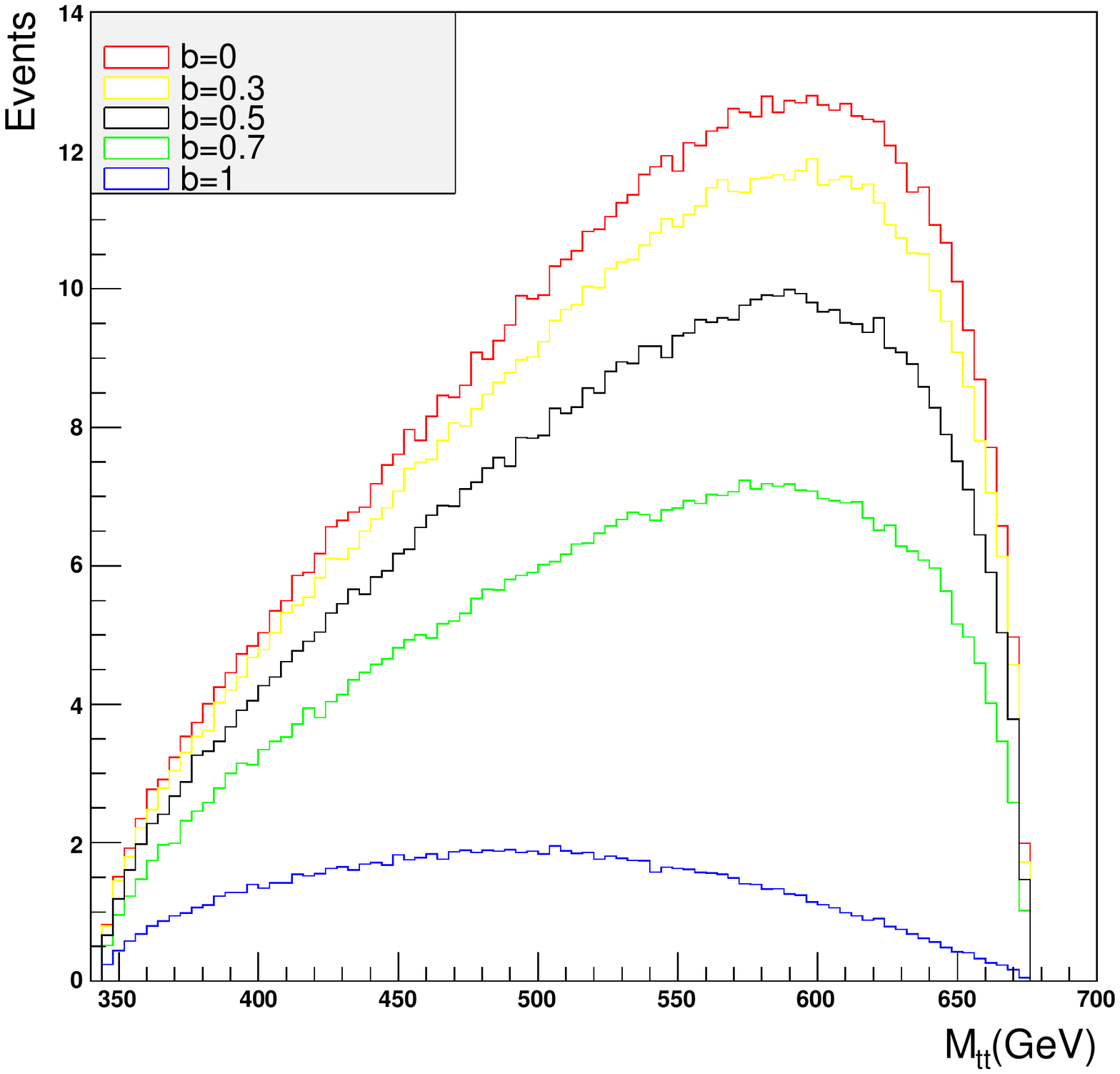}&
\hspace*{-25mm}
\includegraphics[angle=0,width=80mm]{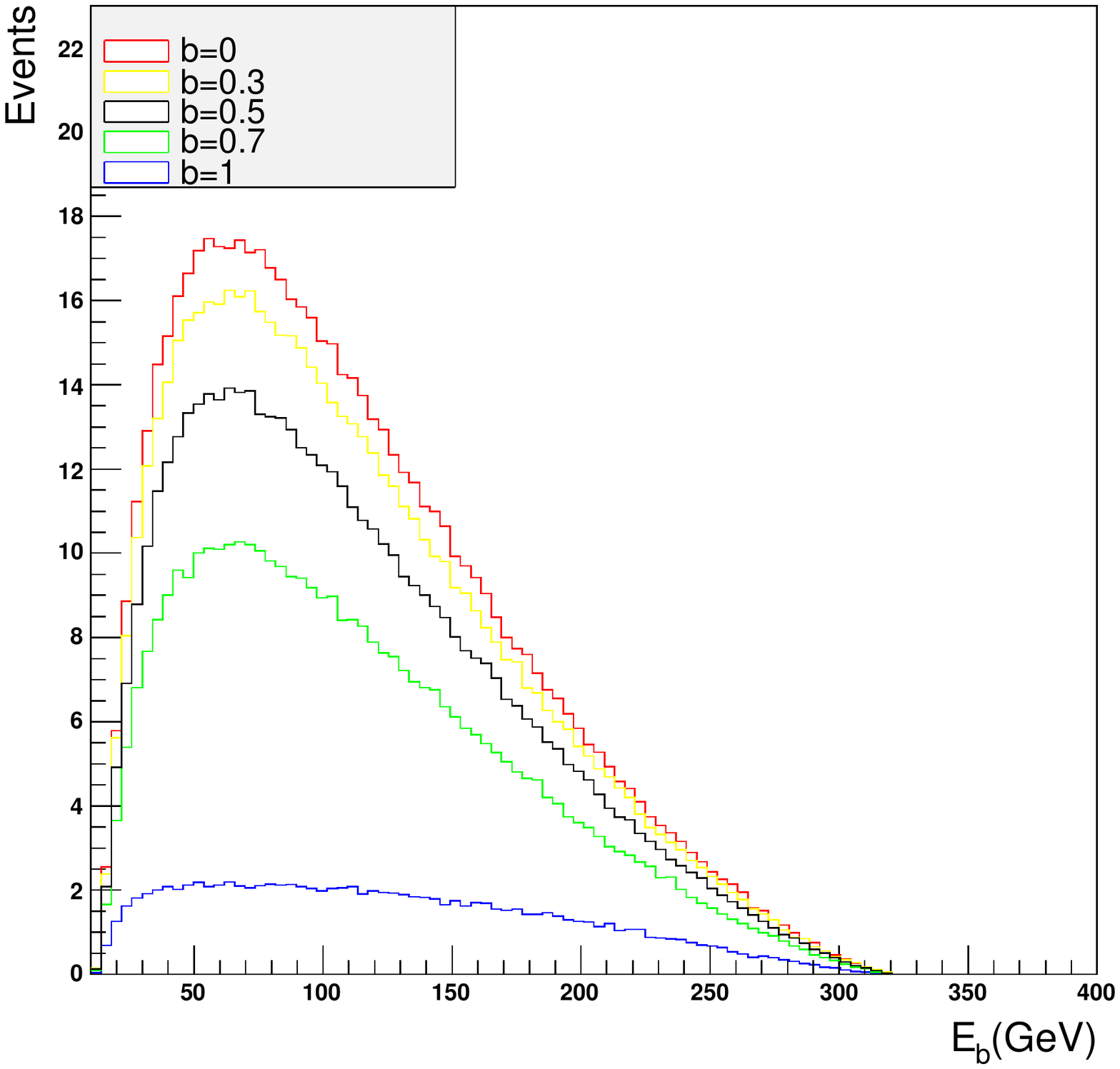} &
\hspace*{-25mm}
\includegraphics[angle=0,width=80mm]{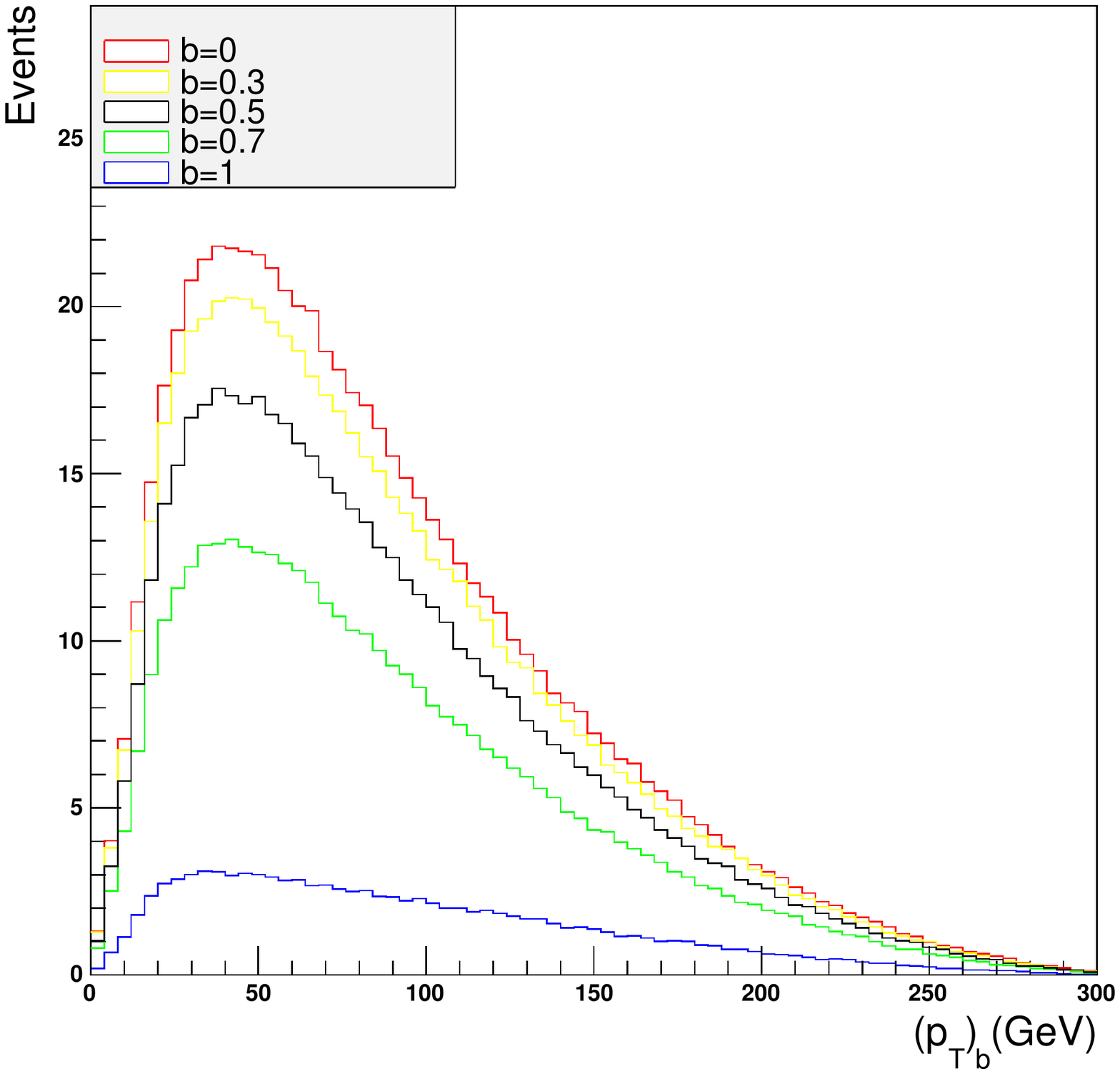}
\end{tabular}
\caption{Model I: Various Distributions for Higgs decay case
at $\sqrt{s}=800$\,GeV with an integrated
luminosity of $300~{\rm fb}^{-1}$ for different $b$ values.}
\label{fig.8}
\end{figure}

Next we come to the impact of CP violation in Higgs Boson decay. As mentioned in
Section~\ref{process} we consider only the $\Phi\rightarrow b\bar b$ decay. 
In Model I, as described in Section~\ref{formalism}, the parameters $a$ and $b$
corresponding to bottom quark are taken to be the same as that corresponding 
to those for the top quark. The signal process we consider is $e^-e^+\rightarrow
t\bar t \Phi\rightarrow t\bar t b\bar b$. As in the previous case, we can contain the 
background by imposing a cut of $124\le M_{b\bar b}\le 126$ on the invariant mass of 
$b\bar b$ ($M_{b\bar b})$. In Fig.~\ref{fig.7} we present the angular, energy 
and $p_T$ distribution of the top quark for the signal process, as well as for the full
process including the background. Comparing the two cases presented, viz the case with no
kinematic cuts (the top row), and the case with cut on the invariant mass, $M_{b\bar
b}$, it is evident that this almost eliminates the background. 
Fig.~\ref{fig.8} shows different distributions of the top quark and the bottom
quark with unpolarized beams. Here again, we see that for large enough values of $b$ it the presence of the
pseudoscalar component can be identified quite easily. As the case with beam
polarization produce similar distributions, we do not displays them here.
The effect of beam polarization is in an
enhancement of the number of events, and thus increase the reach in probing the
value of $b$, as in the case of the process $e^+e^-\rightarrow t\bar t \Phi\rightarrow
b\bar b W^+W^- \Phi$. 

%

\begin{figure}[!t]\centering
\begin{tabular}{c c c} 
\includegraphics[angle=0,width=80mm]{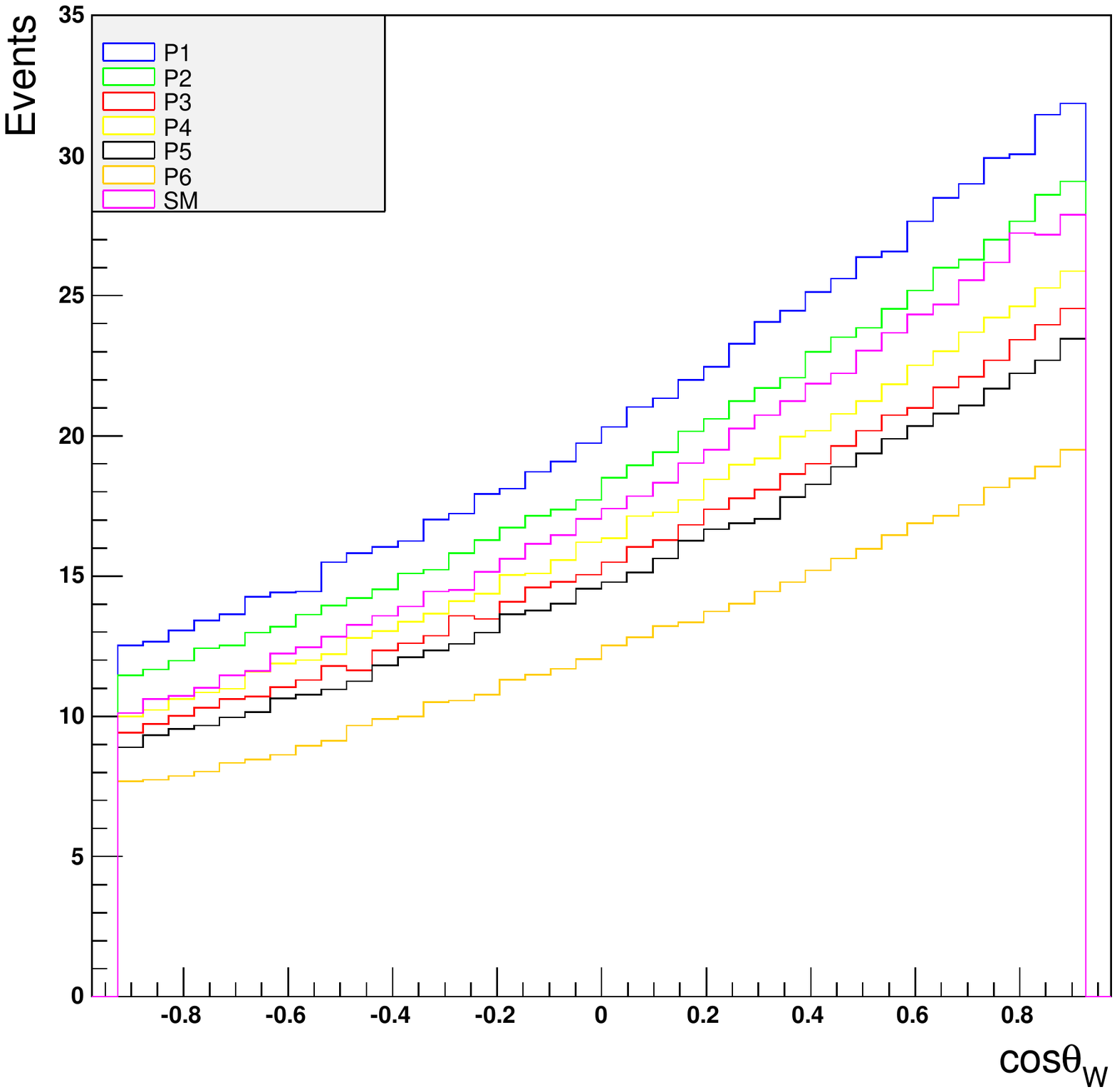} &
\hspace*{-25mm}
\includegraphics[angle=0,width=80mm]{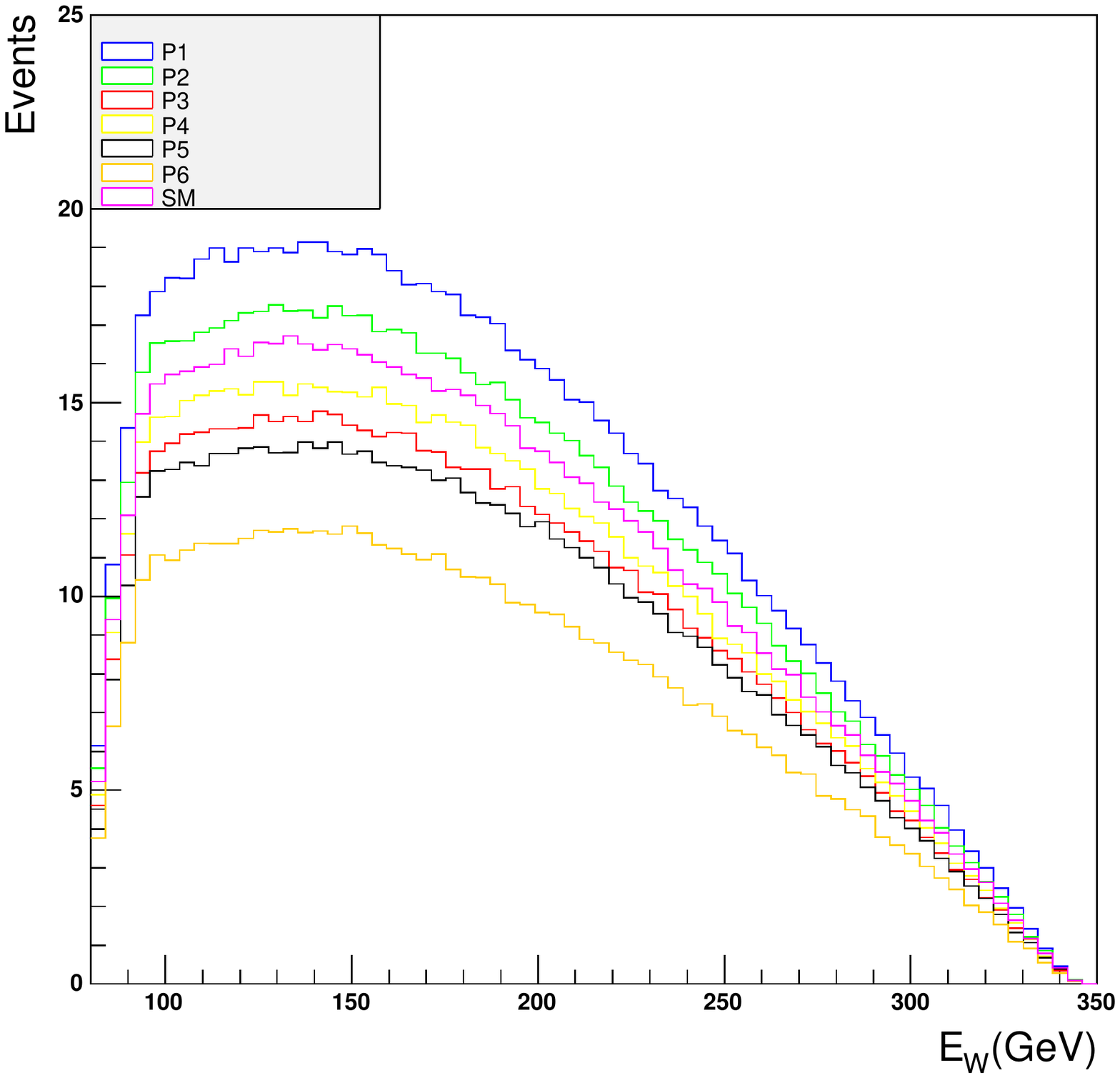} &
\hspace*{-25mm}
\includegraphics[angle=0,width=80mm]{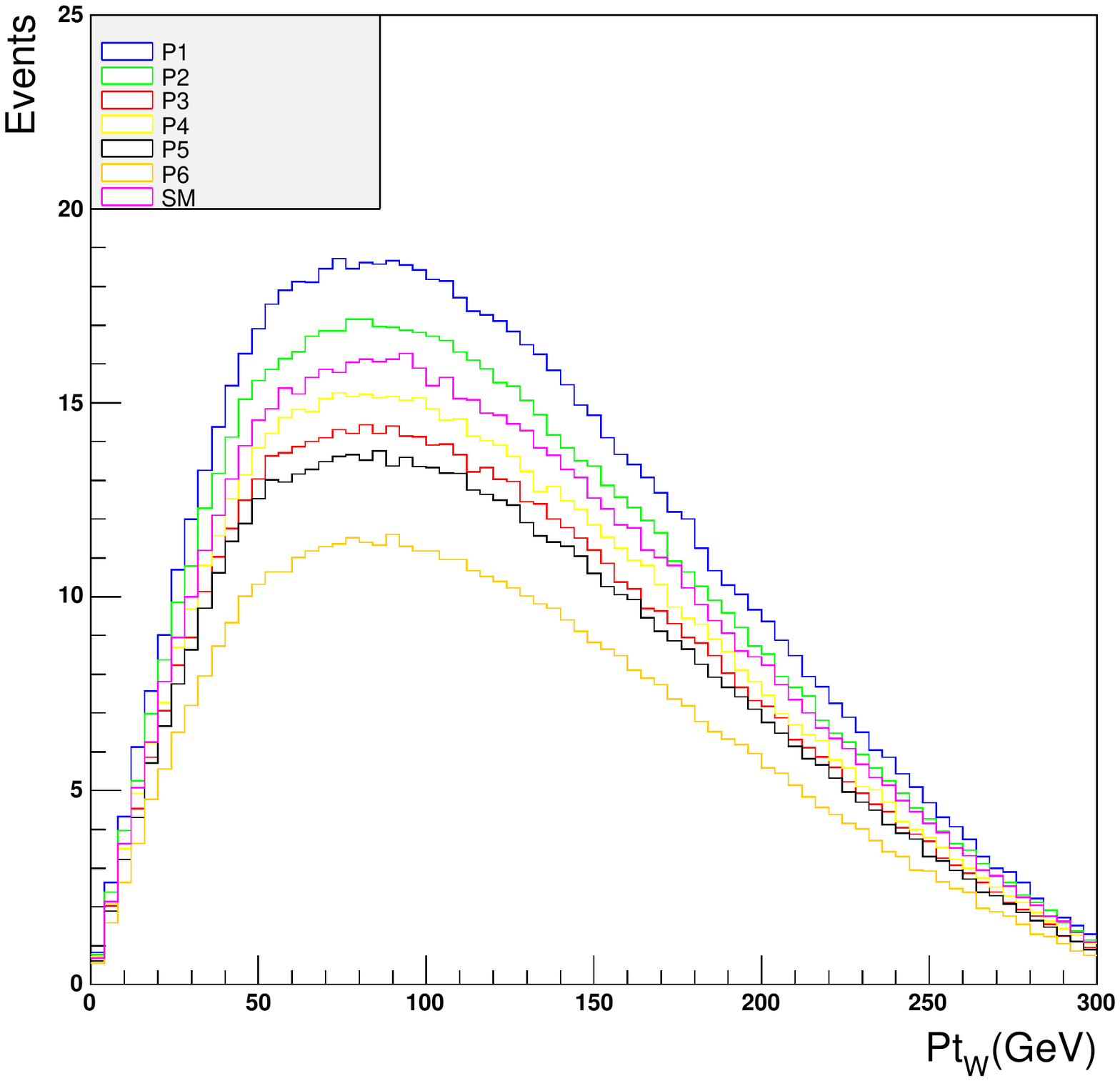}\\
\includegraphics[angle=0,width=80mm]{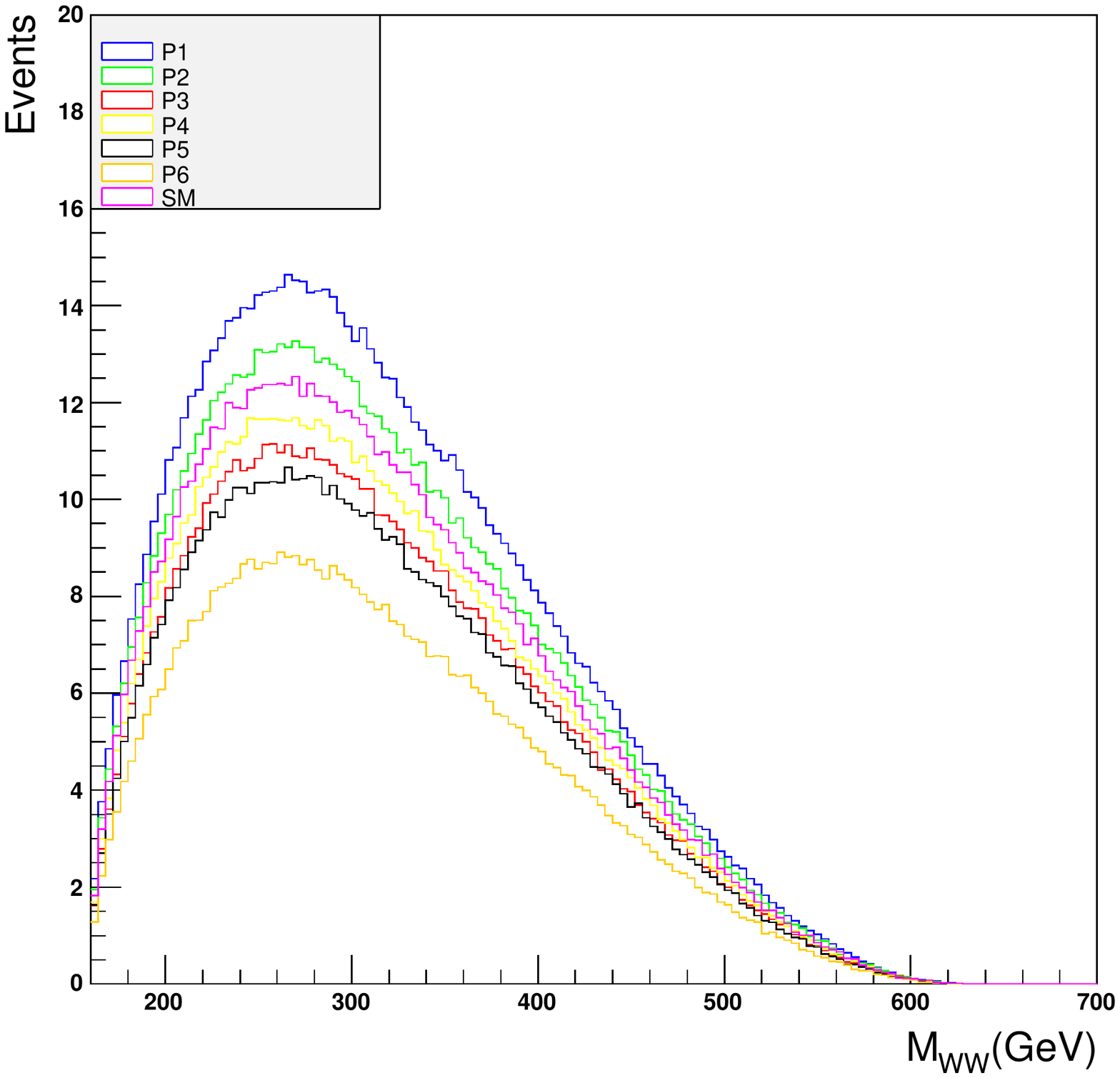} &
\hspace*{-25mm}
\includegraphics[angle=0,width=80mm]{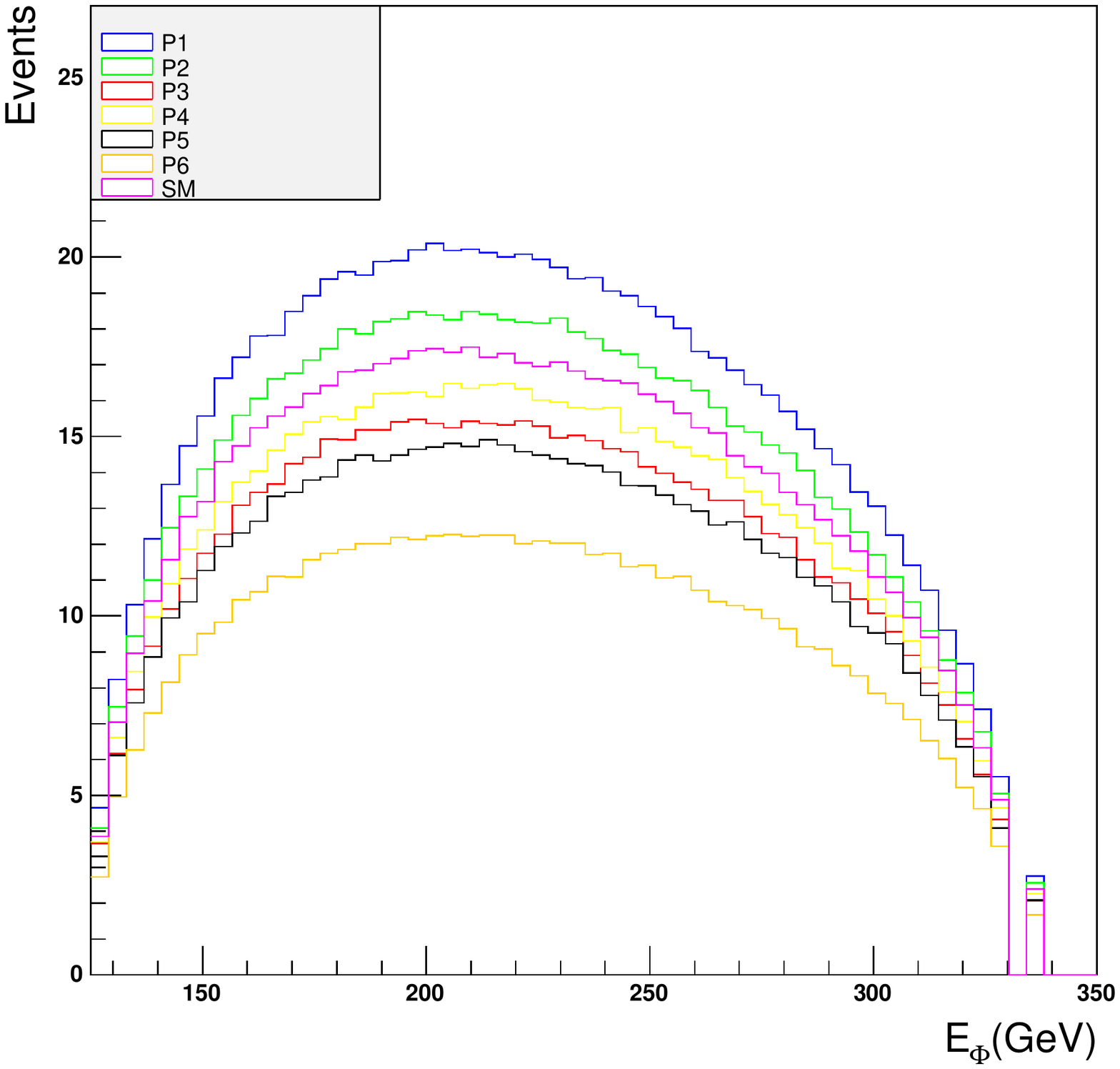} &
\hspace*{-25mm}
\includegraphics[angle=0,width=80mm]{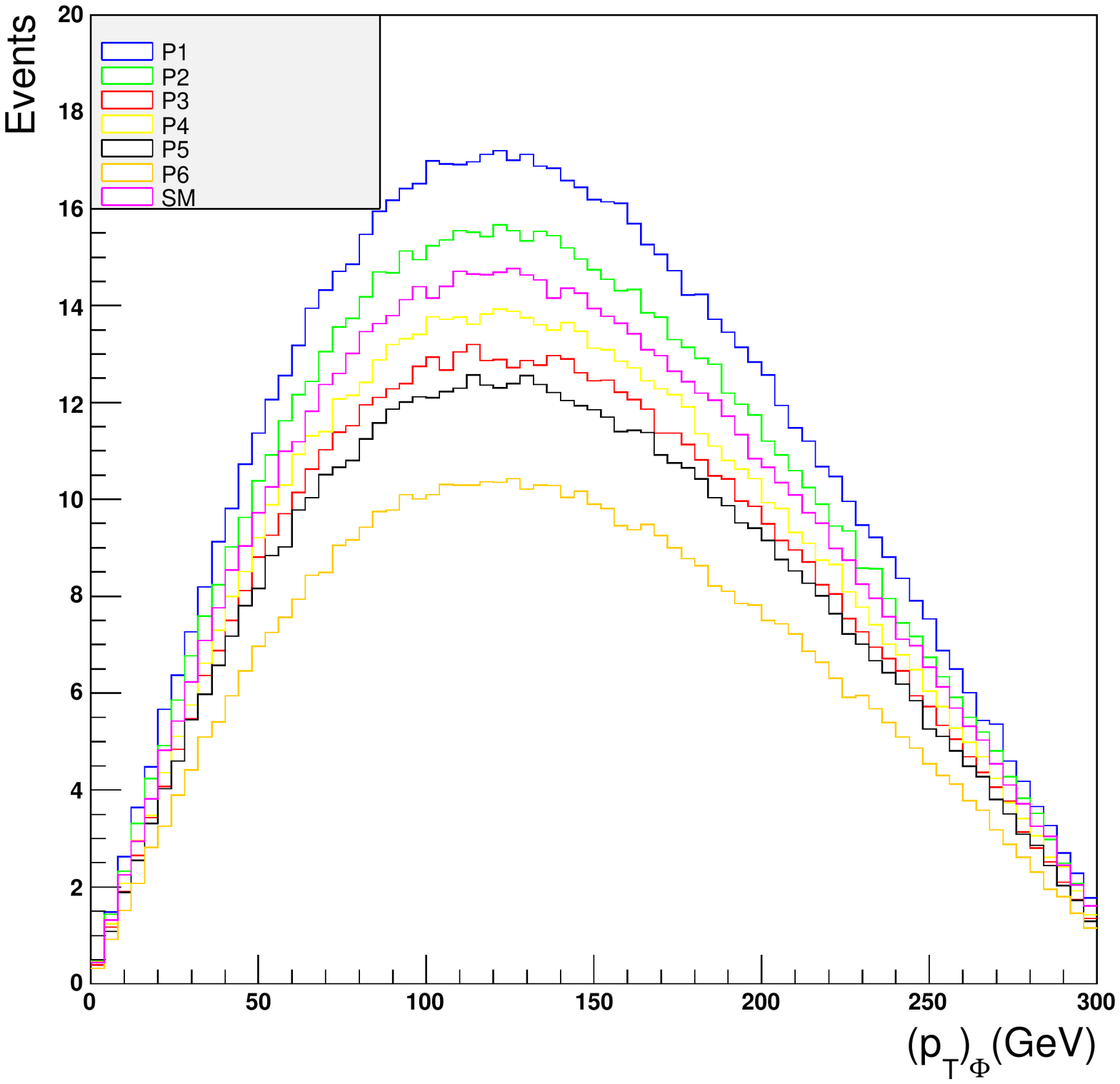}
\end{tabular}
\caption{Model II: Various Distribution in the case of top decay
 at $\sqrt{s}=800$\,GeV with an integrated
luminosity of $300~{\rm fb}^{-1}$ for the  signal
process.} 
\label{fig:MII_up_t_800}
\end{figure}

\begin{figure}[!t]\centering
\begin{tabular}{c c c} 
\includegraphics[angle=0,width=80mm]{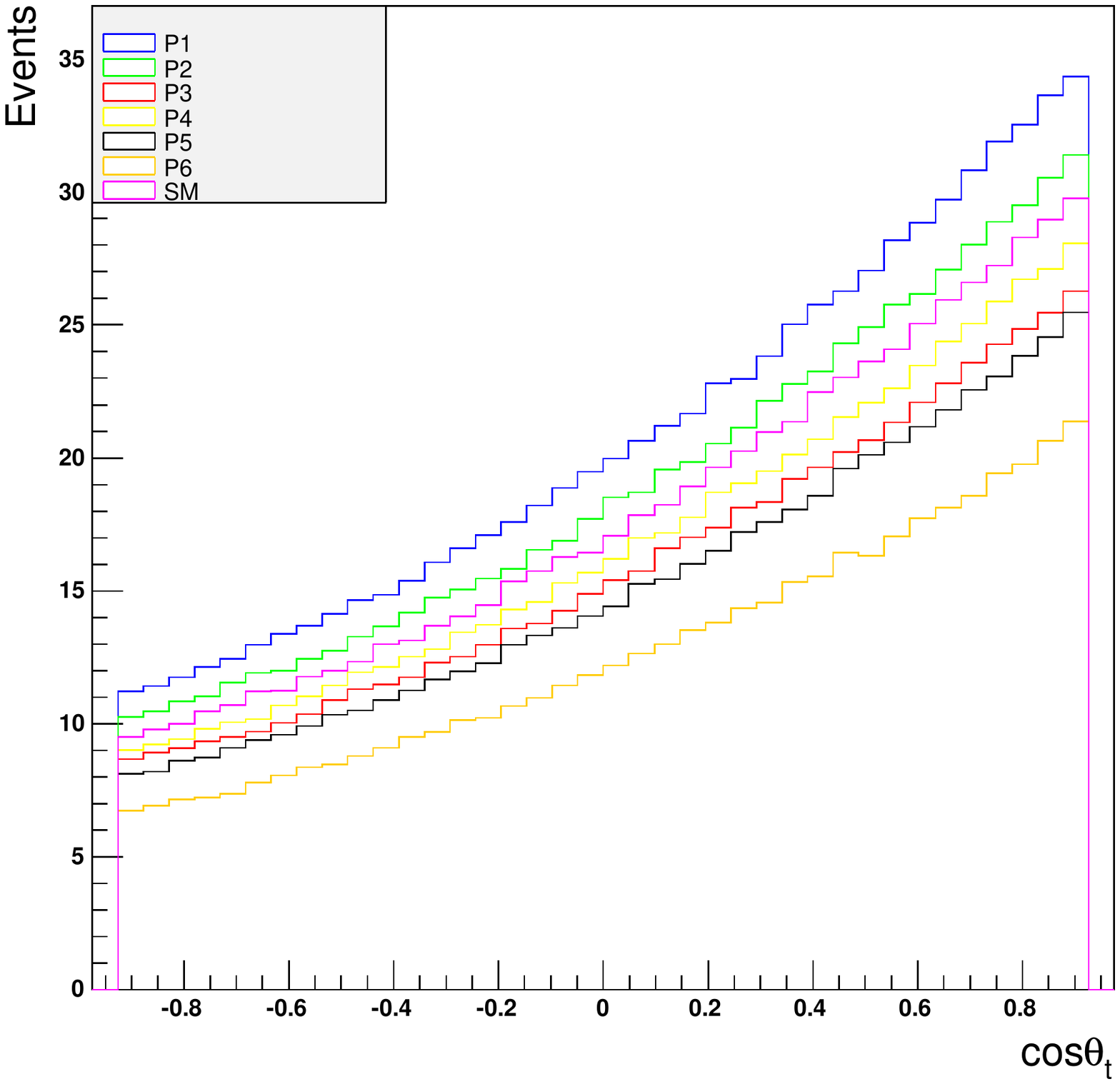}&
\hspace*{-25mm}
\includegraphics[angle=0,width=80mm]{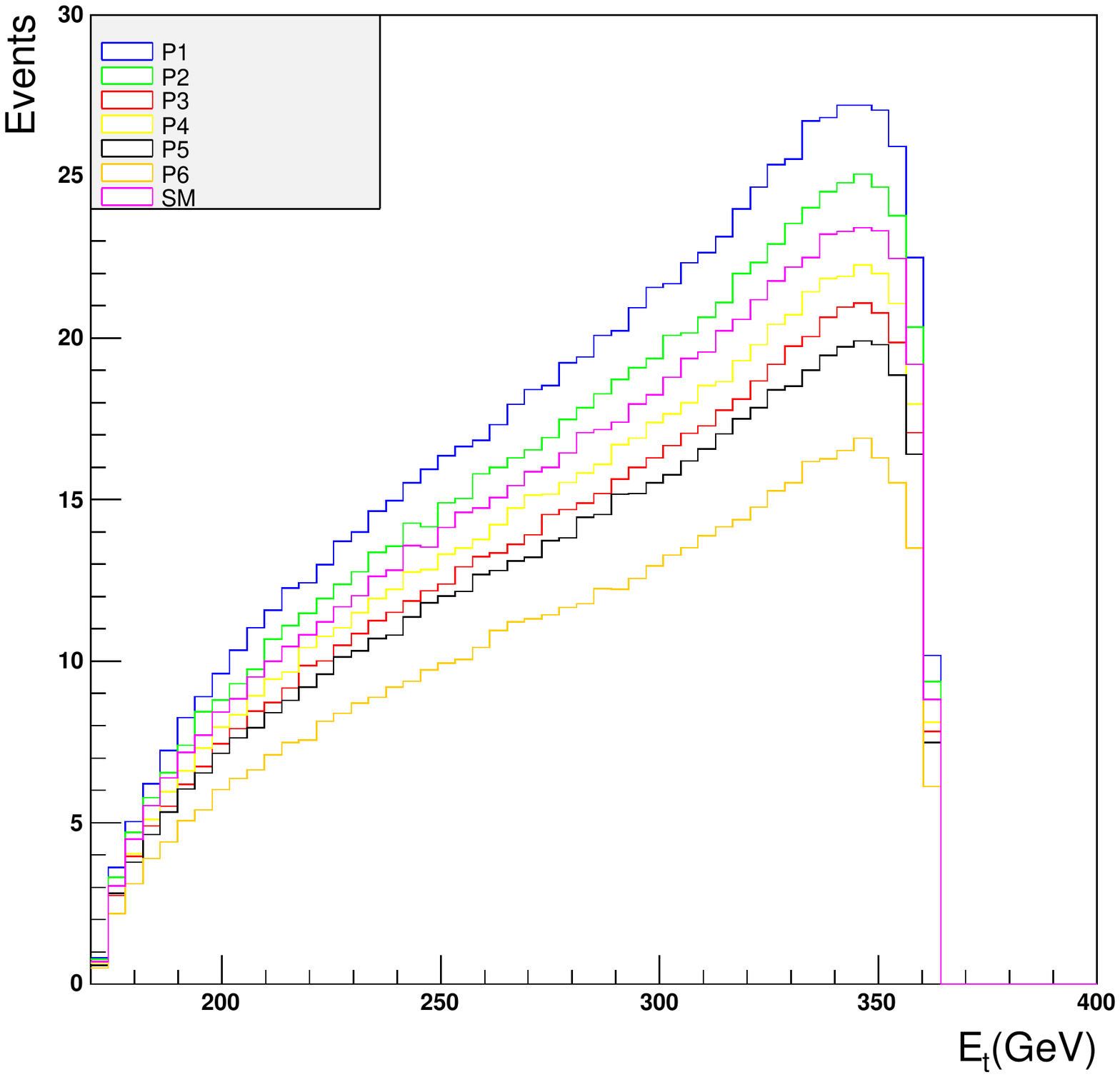} &
\hspace*{-25mm}
\includegraphics[angle=0,width=80mm]{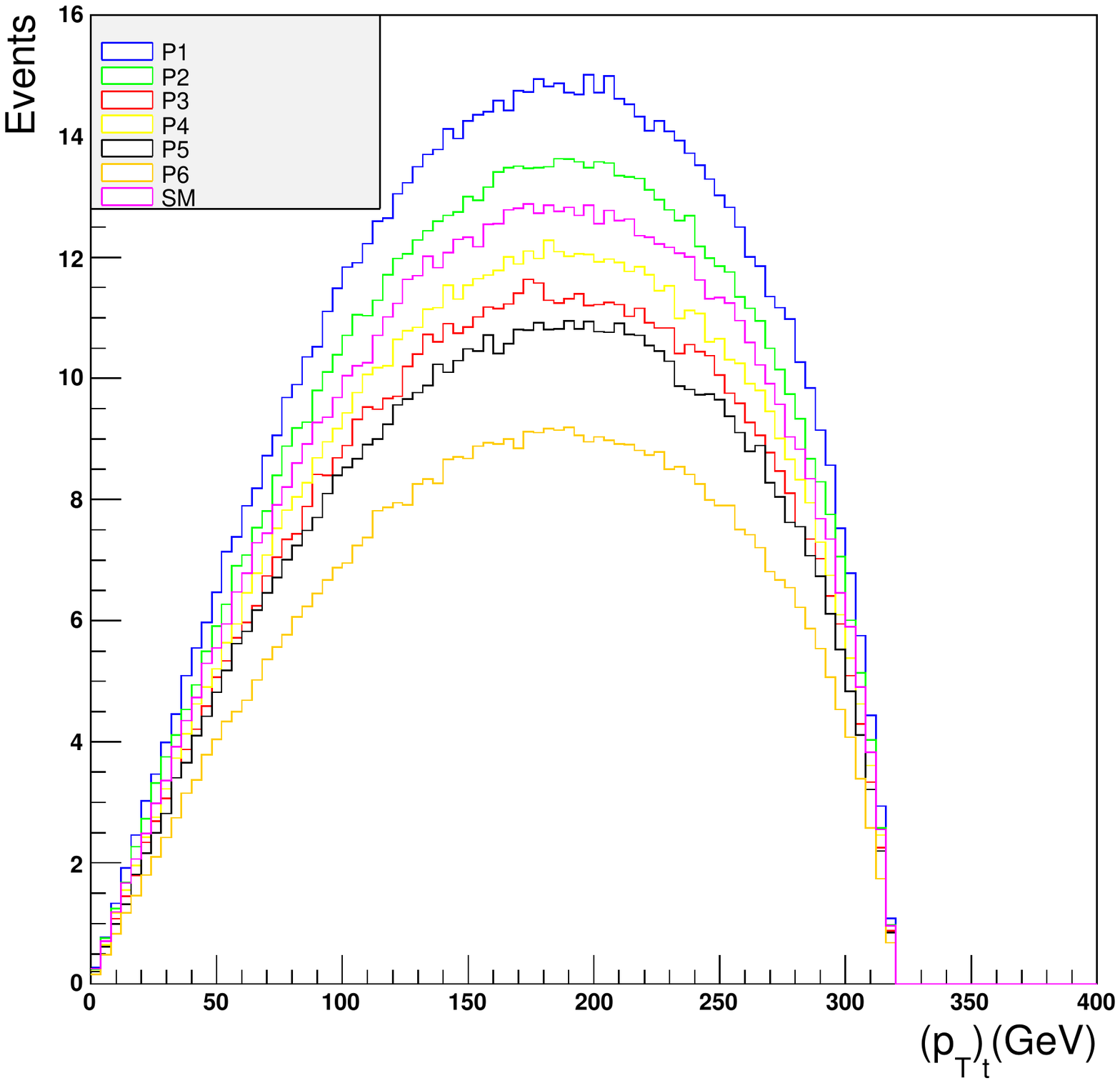}\\
\includegraphics[angle=0,width=80mm]{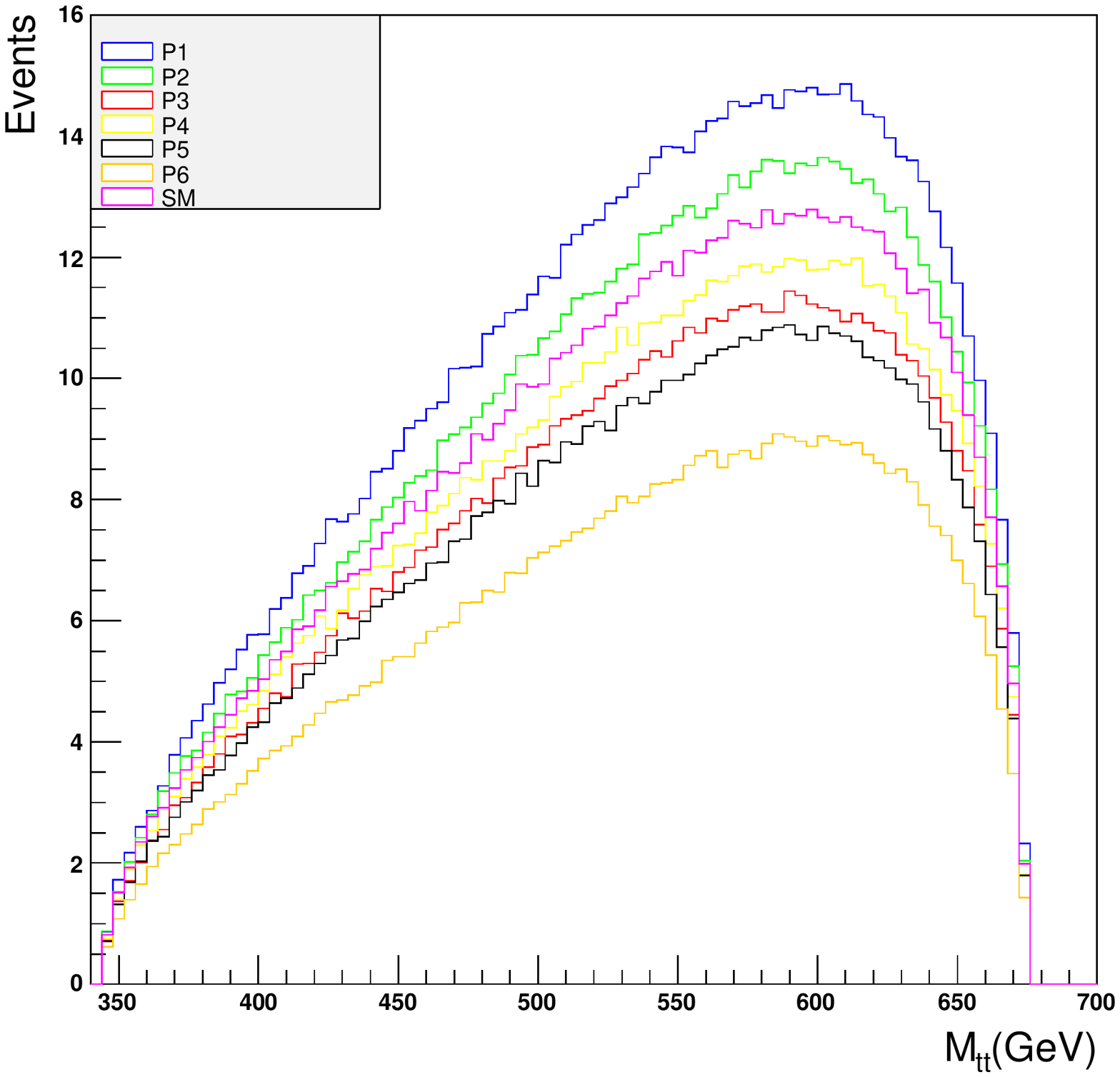}&
\hspace*{-25mm}
\includegraphics[angle=0,width=80mm]{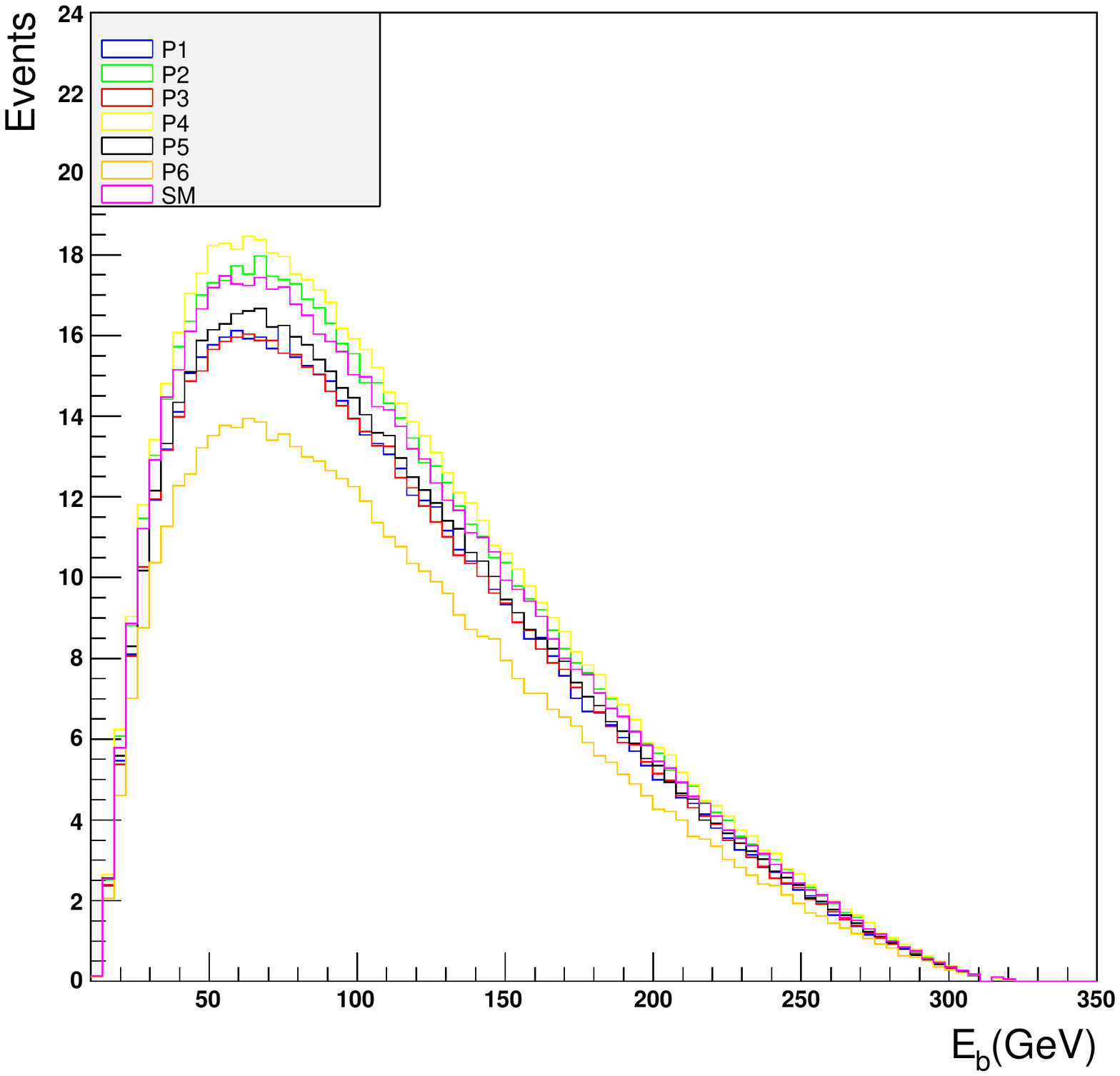} &
\hspace*{-25mm}
\includegraphics[angle=0,width=80mm]{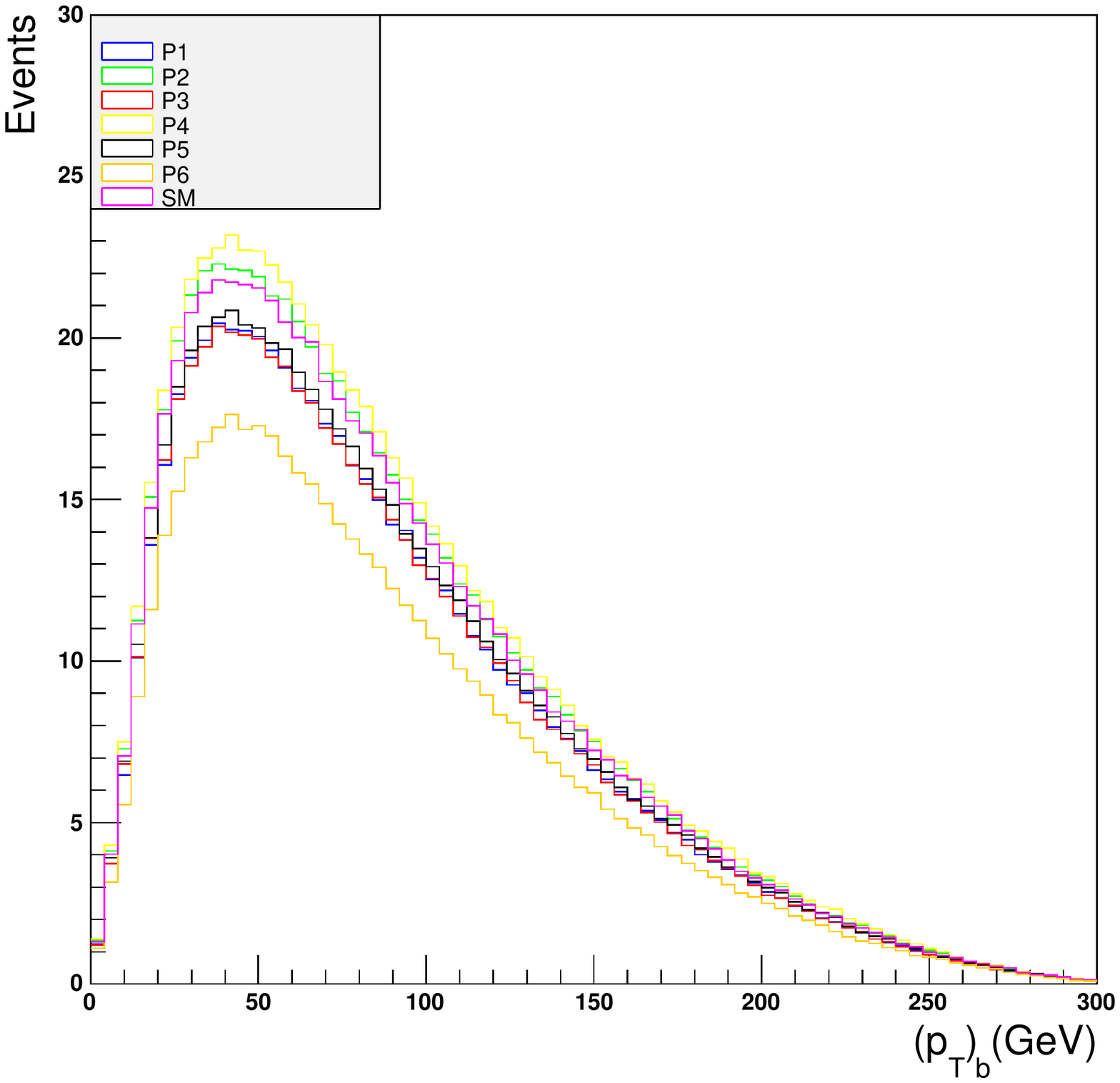} 
\end{tabular}
\caption{Model II: Various Distribution in the case of Higgs decay
 at $\sqrt{s}=800$\,GeV with an integrated
luminosity of $300~{\rm fb}^{-1}$ for the  signal
process. }
\label{fig:MII_up_H_800}
\end{figure}

We next come to the Model II, which is a generic 2HDM with CP violation in the Higgs
sector. As discussed in Section~\ref{formalism}, the parameters, $a$, $b$ and $c$ depend
on the mixing matrix elements ${\cal O}_{ij}$ and $\tan\beta$. For our analysis we
consider only one light Higgs Boson $\Phi=H_1$ of mass around 125\,GeV/$c^2$, with
the other two are heavy enough to be safely away from LHC bounds. Noting that for the
top quark, the parameter $b$ is proportional to $\cot\beta$, we confine our studies to low
$\tan\beta$ cases. Some illustrative values of mixing and the corresponding values of the
couplings are given in Table~\ref{tab:2HDM_coupl}. Our analysis based on these couplings
is given below.

In Fig.~\ref{fig:cs_roots}(middle) we plot the total production cross section for different
mentioned cases. The cross section differs substantially from SM as now $a$ and $b$ parameters
are bit free unlike Model I case where they are constrained by Eq.~\ref{model1constr}. Fig.~\ref{fig:MII_up_t_800} presents the angular, energy and $p_T$ distributions of
the $W^+$, the invariant mass distribution of the $W$ pair, and the energy and $p_T$
distributions of the Higgs Boson in the process $e^+e^-\rightarrow t\bar t
\Phi\rightarrow b\bar b W^+W^- \Phi$. Remember that in case of P1 and P4 we have only
1\% of the pseudoscalar admixture in the Higgs Boson. Notice that unlike in the case of 
Model I, here, the parameters $a$ and $b$ can be larger than 1, and vary with
$\tan\beta$ for the same scalar - pseudoscalar composition of the Higgs Boson.
In P1 case, the deviation from the SM is large, which could further be enhanced with 
the use of beam polarization. For larger $\tan\beta$ value, as indicated by P4, the
effect is negligible for small CP-mixing. 
somewhat larger deviation. With 10\% and 25\% CP-mixing, deviations are significantly
large to be observed in the case of $\tan\beta=20$ (P5 and P6, respectively), 
while for $\tan\beta=2$ (P2 and P3, respectively) the deviations are not so much.
Recall that Model I required much larger mixing for the deviations to be significant. We also presented
the forward-backward asymmetery values in Table~\ref{tab:AFB_M2} for discussed sets. The asymmetry
value show around 7\% deviation from SM in unpolarized case  while for polarized case the deviation is almost
negligible. Coming to the process $e^+e^-\rightarrow t\bar t \Phi\rightarrow t\bar t b\bar b$, the
energy, angle and $p_T$ distributions of the top quark, the invariant mass of the
$t\bar t$ pair, and the energy and $p_T$ distributions of the bottom quark are
presented in Fig.~\ref{fig:MII_up_H_800}. Here again the picture is similar to the
previous case with possibility of large deviations from the SM case even in the case
of small pseudoscalar admixtures.

Clearly, in the case of Model II, even for small pseudoscalar component of the Higgs
Boson,  there can be significant deviation in distributions from their SM values. 
While the analysis of the Higgs decay does not bring out any new features in the
simple considerations of the distributions, it should be possible to construct
specific observables which violate CP symmetry with the help of these final state
particles, and could be the topic of a future study.

\section{Summary and Conclusions}\label{discussion}
In this work motivated by the observation of new scalar resonance\cite{cms, atlas} at LHC, we discussed the
implications of  indefinite CP properties of newly observed state  at proposed linear collider ILC\cite{ILC1, ILC2}. ILC is a next generation $e^--e^+$ collider machine which
apart from searching beyond Standard Model Physics, promise to precisely determine the properties of various SM fields including Higgs at an unprecedented 
accuracy level  which are beyond the realm of currently running LHC. \\

Here we thoroughly investigated the process $e^+e^-\to t \bar{t}\Phi$ with Higgs field $\Phi$ in an indefinite CP state.
In general, CP properties of Higgs Yukawa couplings can be parameterized in terms of its  CP-even(denoted by '$a$') and CP-odd($b$) components along with
its gauge couplings parametrized by multiplicative factor $c$. These parameters
obey certain constrained relations($a^2+b^2=1$, $c=a$) in simplistic scenario, while they can vary much more freely in Models like MSSM or 2HDM for
non supersymmetric case. In this work by categorizing these two scenarios as Model I
and Model II for constrained and unconstrained cases, respectively, we studied
their  implications on various detector level observables. Here we incorporated the decays of the produced particles. Thus our observables are not only sensitive to indefinite CP parameters at production
level but also to the ones which appears at subsequent decay vertex.\\


Compared to earlier model independent studies on this topic, we have considered the decay distributions
of the top quark as well as the Higgs boson to analyse the couplings, and therefore
the CP nature of the Higgs Boson under some realistic scenarios. Such analysis of the spin and parity of the decaying
particles does not require a reconstruction of their polarizations.
We have demonstrated that results based on the study of simplified case of Model I
can differ drastically in a more realistic case like 2HDM of Model II. While the
former require large pseudoscalar admixture in the Higgs Boson, 
and thus large CP violation in the Higgs sector, the latter can produce significant
deviation from the SM case even with 10\% or smaller fraction of pseudoscalar
component in the Higgs Boson. 
Thus, we conclude that the indefinite CP properties of the Higgs
as a window to physics beyond the SM can be probed effectively through the process
considered here.

\newpage

\noindent {\bf{Acknowledgements:}} \\
We thank the authors of {\sc WHIZARD} and the FeynRules interface, 
especially J. Reuter and C. Speckner, for very helpful discussions
regarding the implementation of our models.
BA, SKG and JL thank the Dept.of physics, IITG for their hospitality where part of this work was done.
PP acknowledges the support of BRNS, DAE, Government of India (Project No.:
2010/37P/49/BRNS/1446).

\end{document}